\documentclass[%
 reprint,
superscriptaddress,
nofootinbib,
 amsmath,amssymb,
 aps,
prd,
]{revtex4-2}

\usepackage{graphicx}% Include figure files
\usepackage{dcolumn}% Align table columns on decimal point
\usepackage{slashed}
\usepackage{bm}% bold math
\usepackage{amsmath}
\usepackage{mathtools, nccmath}
\usepackage{amssymb}
\usepackage{amsfonts}
\usepackage{color,soul}
\usepackage[normalem]{ulem}

\begin{document}

\preprint{APS/123}

\title{Self-gravitating nonlinear Dirac fields}% Force line breaks with \\

\author{Huan Zhang}
\affiliation{Institute for Advanced Study, Shenzhen University, Shenzhen, 518000, China}
\affiliation{College of Physics and Optoelectronic Engineering, Shenzhen University, 518060 Shenzhen, China}
\author{Jingsong He}%
\email{Corresponding author: hejingsong@szu.edu.cn}
\affiliation{Institute for Advanced Study, Shenzhen University, Shenzhen, 518000, China}
\affiliation{College of Physics and Optoelectronic Engineering, Shenzhen University, 518060 Shenzhen, China}

\date{\today}% It is always \today, today,

\begin{abstract}
We consider the self-gravitating Dirac field with a scalar fermion self-interaction term. For strong enough attractive fermion self-interaction, the maximum Arnowitt-Deser-Misner mass of soliton solutions consisting of two fermions can exceed the limit of noninteracting Dirac stars classically, and the "particle-like" solutions of the system can exhibit multiplicity, multiple valid solutions for a single set of parameters, that is inherent in the nonlinear Dirac field. We also find the mass-scale separation in our system similar to that discussed in the Einstein-Dirac-Higgs system. Interestingly, the system admits parity-violated solutions. The broken parity symmetry can be restored by increasing the central redshift of the solution.
\end{abstract}

\maketitle

%\tableofcontents

\section{Introduction}
The strong self-gravitating matter field, whose dynamics can be affected by its own gravitational field, serves as an interesting object in theoretical physics. Standard perturbation techniques to find a solution fail in the strong gravitation regime, one has to consider the back reaction of the matter field to the spacetime metric. The self-gravitating classical electromagnetic field, the so-called ``\textit{geon}", was first envisioned and investigated by Wheeler in 1955 \cite{Wheeler}. The singularity-free, asymptotically flat geon will oscillate and decay gradually, indicating that it is unstable and not truly stationary \cite{Wheeler, Power}. Brill \textit{et al}. suggested in Refs. \cite{Brill, Anderson} possible gravitational geons, but it was ultimately proved unstable in Ref. \cite{Perry}. The stable stationary scalar field geons, later called ``\textit{boson stars}" (BSs), were first proposed by Kaup \cite{Kaup}. The localized solution exists due to the non-zero mass of the scalar field. The BSs are prevented from gravitational collapse due to the Heisenberg uncertainty principle manifested in the conservation of $U(1)$ Noether charge. BSs with repulsive or attractive self-interaction have been studied in Refs. \cite{Colpi,Chavanis,Chavanis2}. It is found that Newtonian BSs can collapse into black holes for strong enough attractive self-interaction \cite{Chavanis3}.
    
The self-gravitating fermionic fields, the ``fermion stars",  are less investigated and are much more exotic compact objects than the BSs, while the latter can mimic a black hole \cite{Bustillo} or serve as plausible dark matter candidates \cite{Mielke, Britol, Eby}. The pioneer work about neutrino geons was done by Ruffini and Bonazzola \cite{Ruffini}, but did not provide a solution for the Einstein-Dirac equations. Lee and Pang found explicit solutions for fermion soliton stars while including an additional real scalar field acting as a false-vacuum potential \cite{Lee}, and recent work has extended it beyond the thin-wall approximation \cite{Grosso}. Finster \textit{et al.} found explicit numerical spherically symmetric solutions (FSY solitons) of the Einstein-Dirac equations for the first time \cite{Finster}, where two neutral fermions with opposite spins occupy the shell due to the Pauli exclusion principle. The later works extend to couple the fermionic field with the gauge field \cite{Finster2, Finster3, Dzhunushaliev, Dzhunushaliev2} and Higgs field \cite{Leith}. The inclusion of repulsive interaction of fermions can increase the maximum Arnowitt-Deser-Misner (ADM) mass of Dirac stars as shown in Ref. \cite{Dzhunushaliev3}, similar to those of BSs \cite{Colpi}. Different mass-radius relations of static and spinning scalar, Dirac, and Proca stars with and without self-interaction are investigated in Refs. \cite{Herdeiro, Herdeiro2, Herdeiro3}. In Ref. \cite{Liang}, Liang \textit{et al.} proposed possible hybrid Dirac-Boson stars. The Einstein-Dirac system also allows wormhole solutions in the presence of electromagnetic or phantom fields \cite{Blazquez, Konoplya, Hao}. The fermionic field in the Einstein-Dirac system can also be canonically quantized using the semiclassical gravity approximation \cite{Kain}, in which one can find additional multi-state solutions that are absent in Ref. \cite{Finster}. The massless Einstein-Dirac equations admit infinite redshift solutions \cite{Bakucz, Blazquez2}, explaining the mass-energy spiral in FSY solitons. The fermion ground states of the Einstein-Dirac system become self-trapped at high redshift regions, which becomes more evident when the system has many fermions \cite{Leith2}. The corresponding excited state energy spectra show multivaluedness in forming a loop if the fermion number is larger than four \cite{Leith3}, accompanied by the multiplicity of excited states. The physical mechanism of this energy spectra multivaluedness can be partially explained by the appearance of fermion nodes in the power law zone and the self-trapping of fermions at high redshift regions. The existence of ground state multiplicity of self-gravitating Dirac field has been found recently in Ref. \cite{Leith}, where the fermionic fields are coupled with the gravity and the Higgs field, but the ground state energy spectra are always single-valued \cite{Leith4}. 

In this article, we consider the Einstein-Dirac system with an additional scalar self-interaction of the type $\lambda\bar\Psi\Psi$ between fermions similar to Ref. \cite{Dzhunushaliev3}, where $\lambda>0$ ($\lambda<0$) represents the attractive (repulsive) self-interaction. Here, we focus on the case of attractive interaction. This model can serve as an effective field model for the Einstein-Dirac-Higgs system and has no difficulty generating solutions of low central redshift \cite{Leith4}. We investigate the behaviors of spherically symmetric ``particle-like'' solutions that consist of neutral fermions occupying a filled shell. First, we find that a strong enough attractive self-interaction can lead to the emergence of multiplicity for the ground and excited states. This multiplicity only exists in the unscaled energy spectra and will disappear after rescaling. The rescaled energy spectra remain continuous and single-valued, similar to the results in \cite{Leith, Leith4}. The multiplicity of solutions for our system is inherent in the nonlinear Dirac equations as mentioned in \cite{Finkelstein}. Thus, it has different physical implications from those in Ref. \cite{Leith3}. Second, we find that the maximum ADM mass increases as $\lambda$ increases for large positive values of $\lambda$ and exceeds the limit for noninteracting Dirac stars classically. While the maximum ADM mass decreases as $\lambda$ increases for small positive values of $\lambda$, which is why Ref. \cite{Dzhunushaliev3} did not investigate the case of attractive self-interaction. Nevertheless, we found it follows a similar power law scaling with large values of $\lambda$ as that discovered in Ref. \cite{Dzhunushaliev3}. Third, and more interestingly, in the case of $\lambda>0$, we can only obtain the ground state with the same parity for some range of central redshift, indicating the spontaneous parity symmetry breaking. This symmetry breaking is due to the local negative effective mass of the fermionic fields and can be restored in the high redshift regime.

The paper is structured as follows. In Sec. \ref{sec:II}, we introduce our model of self-gravitating nonlinear Dirac field and derive equations of motion. In Sec. \ref{sec:III}, we introduce the numerical method to find normalized solutions. In Sec. \ref{sec:IV}, we discuss the appearance of multivaluedness in our system. In Sec. \ref{sec:V}, we present our numerical results and analyze the behaviors of different solutions in different parameter settings. Sec. \ref{sec:VI} is our conclusion and discussions.
\section{model set up}\label{sec:II}
We consider a nonlinear Dirac field minimally coupled with Einstein's gravity and work with Planck units $\hbar=c=G=1$, where the Planck length $l_p=\sqrt{\hbar G/c^3}$ and the Planck mass $m_p=\sqrt{\hbar c/G}$. In this article, we adopt the most negative metric signature ($+---$). The corresponding action of the system can be formulated as follows,
\begin{equation}\label{action}
    S = \int\Big(\frac{1}{16\pi}R+\mathcal{L}_d\Big)\sqrt{-g}\,{\rm d}^4x.
\end{equation}
where $\mathcal{L}_d=\bar\Psi(\slashed{D}-m)\Psi+\frac{\lambda}{2}(\bar\Psi\Psi)^2$ is the Lagrangian of nonlinear Dirac field, the first term of $\mathcal{L}_d$ represents the free Dirac field minimally coupled with gravity, the second term represents the self-interaction of the fermions, where $\lambda$ denotes the strength of the interaction. The positive (negative) $\lambda$ represents the attractive (repulsive) interaction. We choose the simplest form of nonlinear self-interaction as in Ref. \cite{Dzhunushaliev} for convenience. $\slashed{D}=i\gamma^\mu D_\mu =i\gamma^\mu(\partial_\mu+\omega_{\mu ab}\hat\gamma^a\hat\gamma^b/4)$, where $\gamma^u$ are the generalized Dirac matrices in curved spacetime satisfying $\lbrace\gamma^\mu,\gamma^\nu\rbrace=2g^{\mu\nu}$, $\Gamma_u=\omega_{\mu ab}\hat\gamma^a\hat\gamma^b/4$ are the spin connection matrices, $\hat\gamma^a$ are the Dirac matrices in flat spacetime satisfying $\lbrace\hat\gamma^a,\hat\gamma^b\rbrace=2\eta^{ab}$, $\omega_{\mu ab}$ are spin connection coefficients. We can transform the gamma matrices $\hat\gamma^a$ in flat spacetime into $\gamma^\mu$ using the tetrad field ${e^\mu}_a$. 

We shall find the stationary spherically symmetric solutions of the system. The metric tensor is defined as a diagonal form in spherical coordinates
\begin{equation}
g_{\mu\nu}=diag\lbrace 1/T(r)^2,-1/A(r),-r^2,-r^2\sin^2{\theta} \rbrace,
\end{equation}
where $T(r)$ and $A(r)$ are real functions of radial coordinate $r$ only.We choose the same tetrad field as in Ref. \cite{Finster}
\begin{equation}
{e^\mu}_a=
\begin{pmatrix}
T & 0 & 0 & 0\\
\sqrt{A}\cos\theta & \sqrt{A}\sin\theta\cos\varphi & \sqrt{A}\sin\theta\cos\varphi & 0\\
-\sin\theta/r & \cos\theta\cos\varphi/r & \cos\theta\cos\varphi/r & 0\\
0 & -\sin\varphi/\sin\theta & \cos\varphi/\sin\theta & 0
\end{pmatrix}.
\end{equation}
The spin connection coefficients $\omega_{\mu ab}$ can be computed conveniently by the formula
\begin{equation}
\omega_{\mu ab} = \frac{1}{2} e_{\mu}^{c} (\lambda_{abc} + \lambda_{cab} - \lambda_{bca}),
\end{equation}
where $\lambda_{abc} = e_{a}^{\mu} (\partial_{\nu} e_{b\mu} - \partial_{\mu} e_{b\nu}) e_{c}^{\nu}$.

The equation of motion of the system can be derived by variation of the action \eqref{action} with respect to Dirac field $\Psi$ and the metric $g_{\mu\nu}$,
\begin{subequations}\label{eom}
\begin{eqnarray}
    \slashed{D}\Psi-m\Psi=-\lambda(\bar\Psi\Psi)\Psi,\\
    R_{\mu\nu}-\frac{1}{2}g_{\mu\nu}R=8\pi T_{\mu\nu},
\end{eqnarray}
\end{subequations}
where the energy-momentum tensor
\begin{equation}
    T_{\mu\nu} = \frac{1}{2}Re(\bar\Psi\,i\gamma_\mu D_\nu\Psi+\bar\Psi\,i\gamma_\nu D_\mu\Psi)-g_{\mu\nu}\mathcal{L}_d.
\end{equation}

 Unlike in Ref. \cite{Dzhunushaliev3}, here we consider $N(\geq2)$ neutral fermions occupying a filled shell with self-interaction. The total angular momentum of $N$ neutral fermions should be zero to maintain the system's spherical symmetry. We can describe the wavefunction of fermions as the Hartree-Fock state \cite{Finster4}
\begin{equation}
    \Psi=\Psi_{j,k=-j}^\pm\,\wedge\Psi_{j,k=-j+1}^\pm\,\wedge\,...\wedge\Psi_{j,k=j}^\pm,
\end{equation}
where $\Psi_{j,k}^\pm$ are wavefunctions with different parities, $j=\lbrace\frac{1}{2},\frac{3}{2},\frac{5}{2},...\rbrace$ are half integers representing the total angular momentum of a fermion. $k$ is the projection of the total angular momentum on the $z$-axis, and the fermion number $N$ is an even number satisfying $N=2j+1$. $\Psi_{j,k}^+$ is the wavefunction with even (odd) parity if $N/2$ is even (odd). While $\Psi_{j,k}^-$ is the wavefunction with odd (even) parity if $N/2$ is even (odd) \cite{Finster4}.
 
We use the fermionic field ansatz similar to that in \cite{Finster4} and assume the wavefunction of each fermionic field $\Psi_{j,k}^+$ has the following form, analogous to the electron orbit of the hydrogen atom
\begin{eqnarray}\label{ansatz}
\Psi_{j,k}^+ = e^{-i\omega t} \frac{\sqrt{T(r)}}{r} \begin{pmatrix}
\chi_{j-\frac{1}{2}}^{k} \alpha(r) \\
i\chi_{j+\frac{1}{2}}^{k} \beta(r)
\end{pmatrix},
\end{eqnarray}
where $\alpha(r)$ and $\beta(r)$ are real functions to be determined, $\omega$ is the oscillation frequency of the system, $\chi_{j-\frac{1}{2}}^{k}$ are eigenstates of the operator $\pmb{\sigma}\cdot \pmb{L}+1$, where $\pmb{\sigma}$ is the Pauli vector and $\pmb{L}$ represents the orbital angular momentum. The eigenstates $\chi_{j\pm\frac{1}{2}}^{k}$ can be described by superpositions of spherical harmonics $Y^l_m(\theta,\phi)$
\begin{subequations}{\label{eig}}
\begin{eqnarray}
\chi_{j-\frac{1}{2}}^{k} & = & \sqrt{\frac{4\pi(j+k)}{2j}}Y_{j-\frac{1}{2}}^{k-\frac{1}{2}}\begin{pmatrix} 1 \\ 0 \end{pmatrix}\nonumber \\
& &+ \sqrt{\frac{4\pi(j-k)}{2j}}Y_{j-\frac{1}{2}}^{k+\frac{1}{2}}\begin{pmatrix} 0 \\ 1 \end{pmatrix},\\
\chi_{j+\frac{1}{2}}^{k} & = & \sqrt{\frac{4\pi(j+1-k)}{2j+2}}Y_{j+\frac{1}{2}}^{k-\frac{1}{2}}\begin{pmatrix} 1 \\ 0 \end{pmatrix}\nonumber \\
& & - \sqrt{\frac{4\pi(j+1+k)}{2j+2}}Y_{j+\frac{1}{2}}^{k+\frac{1}{2}}\begin{pmatrix} 0 \\ 1 \end{pmatrix}.
\end{eqnarray}
\end{subequations}
The extra $\sqrt{4\pi}$ factors in Eqs.\eqref{eig} are introduced to ensure that the eigenstates  $\chi_{j-\frac{1}{2}}^{k}$ are normalized to fermion number $N$
\begin{equation}
\sum^j_{k=-j} \bar\chi_{j\pm\frac{1}{2}}^{k}(\theta,\phi)\chi_{j\pm\frac{1}{2}}^{k}(\theta,\phi)=2j+1=N.
\end{equation}
Substituting Eq. \eqref{ansatz} into Eqs. \eqref{eom}, we can obtain the following coupled ordinary differential equations (ODEs) of $\alpha(r)$, $\beta(r)$, $A(r)$, and $T(r)$
\begin{subequations}\label{ede}
    \begin{eqnarray}
        \sqrt{A}\alpha^\prime & = & \frac{N\alpha}{2r}-\Big(\omega T+m-\lambda V\Big)\beta,\\
        \sqrt{A}\beta^\prime & = & \Big(\omega T-m+\lambda V\Big)\alpha-\frac{N\beta}{2r},\\
        rA^\prime & = & 1-A-8\pi N\omega T^2(\alpha^2+\beta^2)\nonumber\\
        & & -4\pi r^2\lambda V^2,\\
        2rA\frac{T^\prime}{T} & = & A-1-8\pi NT\sqrt{A}(\alpha\beta^\prime-\beta\alpha^\prime)\nonumber\\
        & & +4\pi r^2\lambda V^2,\\
        V & = & \frac{NT}{r^2}(\alpha^2-\beta^2),
    \end{eqnarray}
\end{subequations}
where the prime denotes the differentiation with respect to the radial coordinate.  The terms $\lambda V$ and $4\pi r^2\lambda V^2$ appearing in Eqs. \eqref{ede} contributes crucial effects to our findings mentioned in the introduction. Eqs. \eqref{ede} are highly nonlinear ODEs describing the self-gravitating fermions with self-interactions, in which it is hard to find an analytical solution. Therefore, we seek numerical solutions in this kind of system. We will introduce the numerical method to solve Eqs. \eqref{ede} in the next section.

\section{Numerical method}\label{sec:III}
We expand the Eqs. \eqref{ede} around small $r$ to find regular solutions of the coupled ODEs \eqref{ede}. We recover the same small $r$ expansion in the first few terms as in Ref. \cite{Leith3} for fermions number $N\geq4$ since the contribution to self-interaction comes from the higher-order terms.
\begin{subequations}
    \begin{eqnarray}
        \alpha(r) & = & \alpha_1 r^{N/2} + \mathcal{O}(r^{N/2+2}),\\
        \beta(r) & = & \frac{1}{N + 1}(\omega T_0 - m)\alpha_1 r^{N/2 + 1} + \mathcal{O}(r^{N/2+3})\,,\\
        T(r) & = & T_0 - \frac{4\pi T_0^2 \alpha_1^2}{N + 1}(2\omega T_0 - m)r^{N} + \mathcal{O}(r^{N+2})\,,\\
        A(r) & = & 1 - 8\pi \omega T_0^2 \alpha_1^2 \frac{N}{N + 1}r^{N} + \mathcal{O}(r^{N+2})\,.
    \end{eqnarray}
\end{subequations}
The case $N=2$ is special as the expansion is related to the strength $\lambda$ of the self-interaction
\begin{subequations}
    \begin{eqnarray}
        \alpha(r) & = & \alpha_1 r + \mathcal{O}(r^3),\\
        \beta(r) & = & \frac{1}{3}(\omega T_0 - m+2\lambda T_0\alpha_1^2)\alpha_1 r^{2} + \mathcal{O}(r^4)\,,\\
         T(r) & = & T_0 - \frac{4}{3}\pi T_0^2 \alpha_1^2 [2\omega T_0 - m\nonumber\\
         & & +6\lambda\alpha_1^2T_0]r^{2} + \mathcal{O}(r^4)\,,\\
         A(r) & = & 1 - \frac{16}3\pi T_0^2 \alpha_1^2 (\omega+\lambda\alpha_1^2)r^{2} +  \mathcal{O}(r^4)\,.
    \end{eqnarray}
\end{subequations}
The asymptotic flat boundary conditions are as follows
\begin{subequations}\label{bc}
    \begin{eqnarray}
        \lim_{r\to\infty}\alpha(r) & = & 0,\\
        \lim_{r\to\infty}\beta(r) & = & 0,\\
        \lim_{r\to\infty}A(r) & = &1-\frac{2M}{r},\\
        \lim_{r\to\infty}T(r) & = &\Big(1-\frac{2M}{r}\Big)^{-\frac{1}{2}},
    \end{eqnarray}
\end{subequations}
where $M$ represents the Arnowitt-Deser-Misner (ADM) mass of multi-fermion soliton solutions. To find solutions to Eqs. \eqref{ede}, we set the mass $m$ of the fermionic field and $T_0$ to one, the initial position of integration $r_0=10^{-5}$, and keep $\alpha_1$ fixed, then shoot for the numerical localized solutions by varying frequency $\omega$. We only need to apply the shooting method to the function $\alpha(r)$, the function $\beta(r)$ will converge automatically.

The Noether charge corresponding to $U(1)$ symmetry of the fermionic field is equal to fermion number $N$ and can be calculated by the integration of the conserved current $\bar\Psi\gamma^\mu\Psi$ that is normal to the spacelike hypersurface $\mathcal{H}$
\begin{equation}\label{nc}
    N=\int_\mathcal{H} \,\bar\Psi\gamma^\mu\Psi\, n_\mu\,\sqrt{-h}\,{\rm d}^3x,
\end{equation}
where $n_\mu=\sqrt{g_{tt}}(1,0,0,0)$ is the future-directed timelike vector normal to the spacelike hypersurface $\mathcal{H}$, and $h$ is the determinant of the induced metric on the hypersurface. Then, we can naturally define the fermion density as
\begin{equation}
n_f=\bar\Psi\gamma^\mu\Psi\, n_\mu=\frac{NT(\alpha^2+\beta^2)}{r^2}.
\end{equation}
Eq. \eqref{nc} leads to the normalization condition of the fermionic field
\begin{equation}
    4\pi \int_{0}^{\infty} (\alpha^2 + \beta^2) \frac{T}{\sqrt{A}} {\rm d}r = 1.
\end{equation}
Similarly, the Komar mass, which is the Noether charge of time translation symmetry, can be calculated by \cite{Wald}
\begin{eqnarray}\label{komar}
    M_K = \int_{\mathcal{H}}\sqrt{-h}\,n_\mu K^\nu(2{T^\mu}_\nu-T{\delta^\mu}_{\nu}){\rm d}^3x\nonumber\\
     =  \int_0^{+\infty}\frac{4\pi N}{\sqrt{A}}[2\omega T(\alpha^2+\beta^2) -m(\alpha^2-\beta^2)]{\rm d}r,
\end{eqnarray}
 where $K^\nu=(1,0,0,0)$ is the timelike Killing vector. The form of Eq. \eqref{komar} demonstrates that terms about self-interaction cancel out and have no explicit contribution to the Komar mass. However, the self-interaction of the fermionic field can contribute to the energy-momentum tensor $T_{\mu\nu}$ and implicitly affect the local property and density distribution of the fermionic field. Additionally, the Komar mass should coincide with the ADM mass in asymptotically flat spacetime for our stationary solutions. Numerically, the difference between the Komar and ADM mass is kept less than $10^{-5}$ to ensure that the obtained solutions are accurate. The radius of the fermions $R$ is defined by enclosing $99.9\%$ of the ADM mass
 \begin{equation}\label{radius}
    \frac{R}{2}[1-A(R)]=0.999M.
 \end{equation}

It is not easy to implement boundary conditions \eqref{bc} directly in the numerical computation. Therefore, we apply the rescaling technique proposed in Ref. \cite{Finster} after obtaining a solution numerically with the following weaker boundary conditions 
 \begin{subequations}
     \begin{eqnarray}
         \lim_{r\to\infty}\Big(1-\frac{2M}{r}\Big)^{\frac{1}{2}}T(r)=\tau<\infty,\\
         \Bigg(4\pi \int_{0}^{\infty} (\alpha^2 + \beta^2) \frac{T}{\sqrt{A}} {\rm d}r\Bigg)^{\frac{1}{2}}=\kappa<\infty.
     \end{eqnarray}
 \end{subequations}
The rescaling relations are as follows
\begin{subequations}
    \begin{eqnarray}
        \tilde{\alpha}(r) & = & \sqrt{\frac{\tau}{\kappa}}\alpha(\kappa r),\\
        \tilde{\beta}(r) & = & \sqrt{\frac{\tau}{\kappa}}\beta(\kappa r),\\
        \tilde{A}(r) & = & A(\kappa r),\\
        \tilde{T}(r) & = & \tau^{-1}T(\kappa r).
    \end{eqnarray}
\end{subequations}
We can verify that rescaled fields $\tilde{\alpha}(r)$,  $\tilde{\beta}(r)$, $\tilde{A}(r)$, and $\tilde{T}(r)$ still satisfy Eqs. \eqref{ede} if we rescale the mass $m$ and energy $\omega$ as follows
\begin{equation}
    \tilde{m}=\kappa m,\,\tilde{\omega}=\kappa\tau\omega.
\end{equation}
Accordingly, the Komar mass \eqref{komar} should be rescaled as well
\begin{gather}
    \tilde{M}_K=\frac{\tau}{\kappa}M_K.
\end{gather}
We can obtain one-parameter family solutions by varying $\alpha_1$ and define the central redshift of the solution $z=T(r_0)-1$ as the physically relevant parameter to study the energy spectra \cite{Bakucz}. In the following sections, we omit the tilde of rescaled quantities when presenting our numerical results for convenience.

As mentioned in Ref. \cite{Leith}, it is computationally consuming to find normalizable solutions of the Einstein-Dirac equations in multivalued regions. Here, we use the ODE solving package DifferentialEquations.jl of \textit{Julia} \cite{Bezanson} to generate solutions efficiently. We use ``Vern9", an explicit 9th order Runge-Kutta method ODE solver with adapted time steps, to solve the Eqs. \eqref{ede} with high numerical accuracy.
\begin{figure*}[bpt]
    \centering
    \includegraphics[width=2\columnwidth]{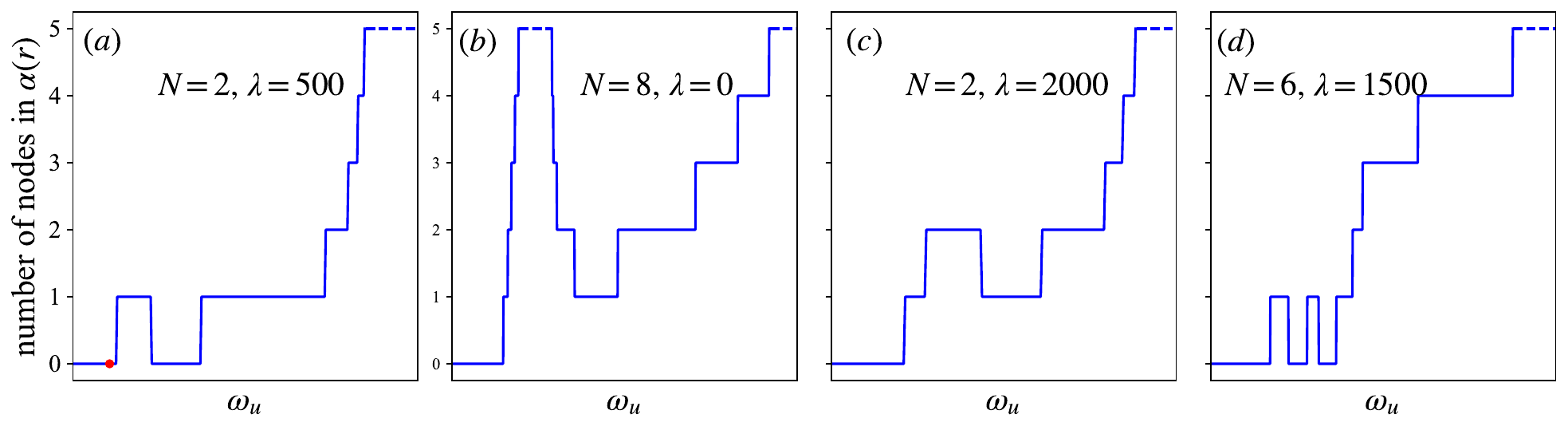}
    \caption{(a)-(d)The number of nodes in $\alpha(r)$ as a function of unscaled energy $\omega_u$ for different numbers of fermions and different strengths of self-interaction. Here, the dashed line in both pictures represents the omitted infinite tower of node numbers. The above figures are only sketches since the lengths of some tiny plateaus are enlarged for better illustration.}
    \label{fig:nodesnumber}
\end{figure*}
\section{Multiplicity of solutions}\label{sec:IV}
As discussed in the previous section and Ref. \cite{Finster}, solutions of the Einstein-Dirac equation can be classified by different parities \cite{Finster2}, with the even (odd) parity denoting the even (odd) sum of node (zero) numbers $n$ in functions $\alpha(r)$ and $\beta(r)$ if $N/2$ is even, and the odd (even) parity denoting the even (odd) sum of node numbers $n$ in functions $\alpha(r)$ and $\beta(r)$ if $N/2$ is odd, e.g., $n=0,2,4,6,...$ for odd parity and $n=1,3,5,7,...$ for even parity if fermion number $N=2$. We can set the unscaled fermion mass $m=\pm1$ to find solutions with different parities numerically. Ref. \cite{Leith3} investigates the fermion states with an even number of nodes. They find that if the number of fermions is larger than four, there are multiple legitimate $n$th excited states, which are characterized by $n$ nodes of the fermion wave functions for a single value of unscaled $\alpha_1$. The multiplicity of solutions persists after rescaling. The rescaled excited energy spectra can form a loop around high redshift regions due to the self-trapping of fermions. The loop might extend to the infinite redshift region with increasing fermion number. 

Nevertheless, some of these states are not literally "excited" since their energies can be lower than the "ground state" (zero node) energy after rescaling in the oscillation regions. However, this classification of different families of solutions is reasonable for the reason that the states with the same node number are continuously connected. For noninteracting Dirac fields coupled with gravity, the emergence of the multivaluedness is controlled by the discrete number $N$. For our system, the unscaled energy spectra can also develop multivaluedness for both the ground state and the excited state for strong enough fermion self-interaction. We should emphasize here that the multiplicity of solutions also exists for the nonlinear Dirac equation in the flat spacetime, which has already been discussed in Ref. \cite{Finkelstein}. The coupling with Einstein's gravity somehow reduces the parameter range that shows multivaluedness. This suggests that the multivaluedness discussed here and in Ref. \cite{Leith3} might have different origins. 

The existence of the multivaluedness suggests that it is impossible to numerically generate all ground state solutions for a single value of $\alpha_1$ by a simple bisection method based on the node number of wavefunctions. Nevertheless, we find that a single binary search can still yield a valid solution, around which we can do a thorough energy sweep to locate the remaining ones. This gives us advantages in generating ground-state solutions efficiently. Fig. \ref{fig:nodesnumber} shows the number of nodes in $\alpha(r)$ as a function of the unscaled energy $\omega$. The normalizable states reside near the step corners, e.g., the location of one of the solutions is shown by the red point in Fig. \ref{fig:nodesnumber}(a). We can see in Fig. \ref{fig:nodesnumber}(a) that there is a multivalued region appearing in the ground state when $N=2$ and $\lambda=500$, as shown by the one-node plateau near the red point. Fig. \ref{fig:nodesnumber}(b) shows a typical energy spectrum for excited states when $N=8$ without self-interaction of fermions, where two infinite towers exist. There is no infinite tower on the one-node plateau in Fig. \ref{fig:nodesnumber}(a), so the excited states cannot reside in this plateau. For strong enough attractive self-interaction, a new two-node plateau featuring the multivaluedness of the excited state will appear in the one-node plateau, as shown in Fig. \ref{fig:nodesnumber}(c).  Fig.  \ref{fig:nodesnumber}(d) exhibits two isolated one-node plateaus, implying a further degree of multivaluedness. We should note here that an infinite tower does not necessarily exist in \textit{multivalued} regions for excited states of noninteracting fermionic fields. However, we cannot find an infinite tower of node numbers for our system. Unlike excited states of noninteracting self-gravitating fermions, the rescaled $\omega-z$ curve of the ground and excited states for $\lambda>0$ is always single-valued and remains continuous for this kind of multivaluedness. More interestingly, as shown in the following sections, the ground state wavefunction $\beta(r)$ will develop a node while the function $\alpha(r)$ remains zero-node, making the previous classification of different solutions inappropriate. This development of the node in the wavefunction is a sign of parity violation since solutions of the system favor only one kind of parity in this case. The transition between different parities is continuous as we tune the parameter $\alpha_1$. Thus, we feel that the node number $n_\alpha$ of the function $\alpha(r)$ can be a better label for different families of states. In this article, we define the ground state as the $n_\alpha=0$ state and the excited states as $n_\alpha=1,2,3,...$ states. In the next section, we will present our numerical results and show that this feature of the wavefunctions is related to the negative effective mass of nonlinear Dirac fields, while the coupling with Einstein's gravity can suppress the parity violation.
\section{Numerical results}\label{sec:V}
In this section, we investigate the behavior of ground states and the excited states of Eqs. \eqref{ede} for different $\lambda$ and fermion number $N$. We begin by studying the ground state for two fermions.
\subsection{$N=2$ and $n_\alpha=0$}
%\subsubsection{Ground states of N = 2 with different $\lambda$}

\begin{figure*}
\centering
\includegraphics[width=2\columnwidth]{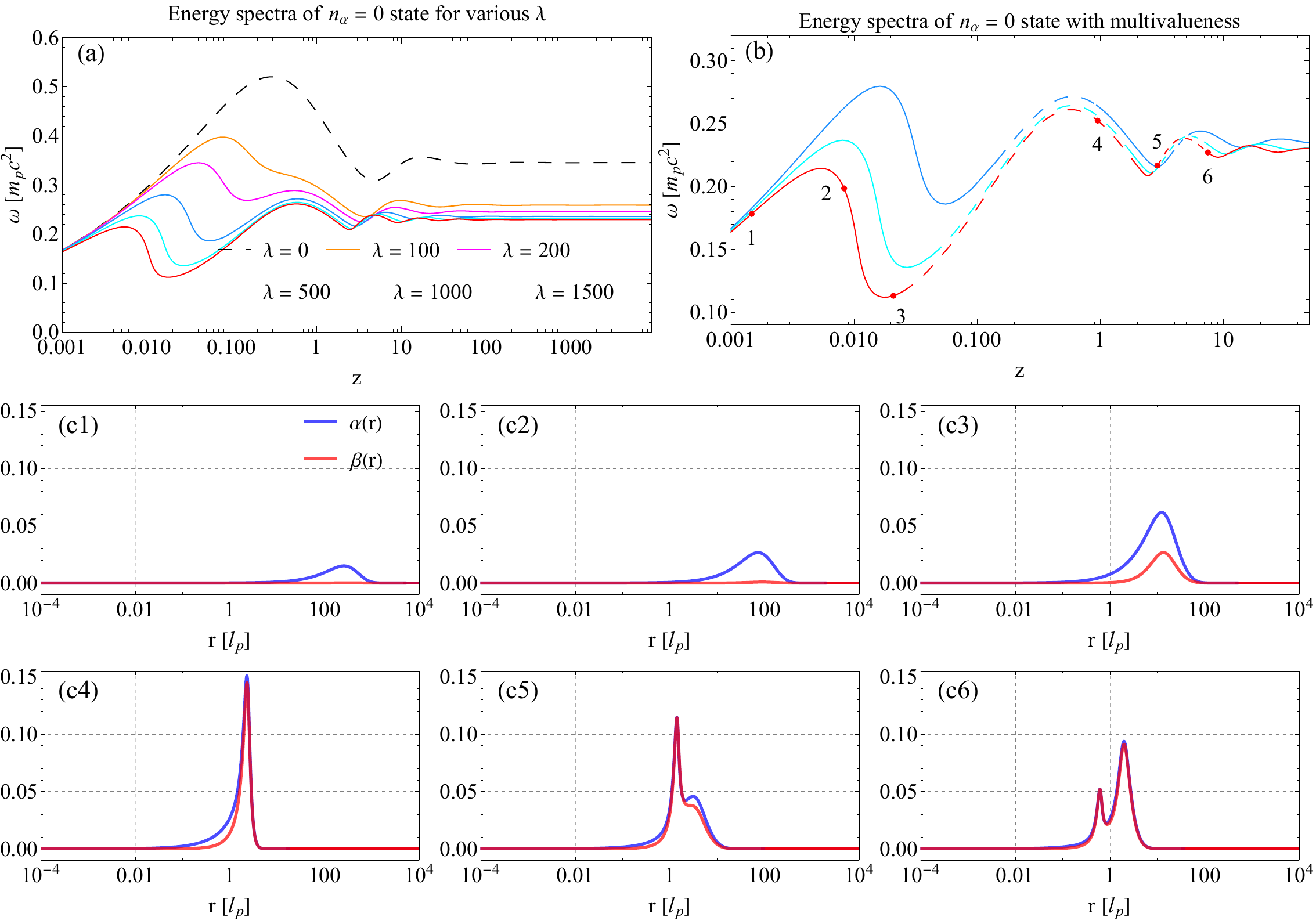}
\caption{(a) The energy of two fermions $\omega$ as the function of the central redshift $z$ with different strength of self-interaction $\lambda$. (b) Replot the energy spectra in (a) that is multivalued for the unscaled energy. The dashed lines in the figure are rescaled from the regions of multivalued energy spectra. (c1)-(c5) The fermion wavefunctions $\alpha(r)$ and $\beta(r)$ at different points of the $\lambda=1500$ energy curve.}
\label{fig:spectra} 
\end{figure*}

Fig. \ref{fig:spectra} shows the energy spectra of the system with different $\lambda$. We can see from Fig. \ref{fig:spectra}(a) that a new trough after the first spectra peak gradually develops with increasing self-interaction and eventually becomes a new minimum in the lower and lower redshift region. Meanwhile, the ground state energies in the low ($z<1$) and high ($z>100)$ redshift regions also decrease as $\lambda$ increases, except for some overlaps in the intermediate regions of oscillation. Furthermore, the ground state energies of the case $\lambda>0$ are lower than that of $\lambda=0$ across all redshifts. It has been proved that $\omega\sim z^{1/4}$ as $z\to0$ for $\lambda=0$ in Ref. \cite{Leith2}, which still holds for $\lambda>0$ but the deviation from the scaling relation (or approximately, the $z<<1$ region of $\lambda=0$ spectrum) at the same redshift becomes larger as $\lambda$ increases. The effect of increasing $\lambda$ in sharpening the other minima of the curve is similar to that of increasing the fermion number in Ref. \cite{Leith3}. We know from the previous section that the unscaled fermion energy spectra for strong enough ($\lambda=500,\,1000,\,1500$ in the case of $N=2$) self-interaction can possess multivaluedness that will not appear after rescaling. Thus, we plot the energy spectra again for strong self-interaction in Fig. \ref{fig:spectra}(b) for better illustration and show the regions rescaled from the multivalued regions by dashed lines. We can see from the figure that the energy spectra is single-valued and continuous. Interestingly, two disconnected regions are denoted by dashed lines lying around the second and third maximum of the spectrum. In Ref. \cite{Leith3}, the authors must construct a "redshift-like" parameter, which requires prior knowledge about the $\omega-z$ curve to transform the energy spectrum into a single-valued curve. Here, the central redshift of the solution automatically becomes an ideal monotonic single-valued parameter to parameterize the energy spectrum without forming a loop. Combining Fig. \ref{fig:spectra}(a) and (b), there must be a threshold for the appearance of multivaluedness existing between $\lambda=200$ and $\lambda=500$. The accurate value of this threshold is not easy to determine since one has to sweep the entire energy curve to find whether the solution shows multiplicity. Ref. \cite{Leith} also discussed the multivaluedness of the ground state for self-gravitating fermionic fields coupled with the Higgs field through the Yukawa term $\mu h\bar\Psi\Psi$\footnote{The discussion here adopts same notations as in Ref. \cite{Leith}.}, where $\mu$ is the couple constant and $h$ is the Higgs field with the "Mexican hat" potential $V(h)=\lambda(h^2-v^2)^2$. The fermions acquire mass $m_f=\mu v$ by the Higgs mechanism and attract each other by exchanging a Higgs boson, which can be understood by the effective attractive Yukawa potential $V_{Y}=-\mu e^{-\gamma m_fr}/r$ at the lowest order, where $\gamma$ is a scaling factor. Ref. \cite{Leith} finds the emergence of multivaluedness for small parameter $v$ and $\xi=2\sqrt{2}\lambda/\mu$ and shows that both smallness of $v$ and $\xi$ imply strong coupling (large $\mu$), which suggests the multivaluedness of our system might have the same origin as that in Einstein-Dirac-Higgs system.

\begin{figure}[bpt]
\centering
\includegraphics[width=\columnwidth]{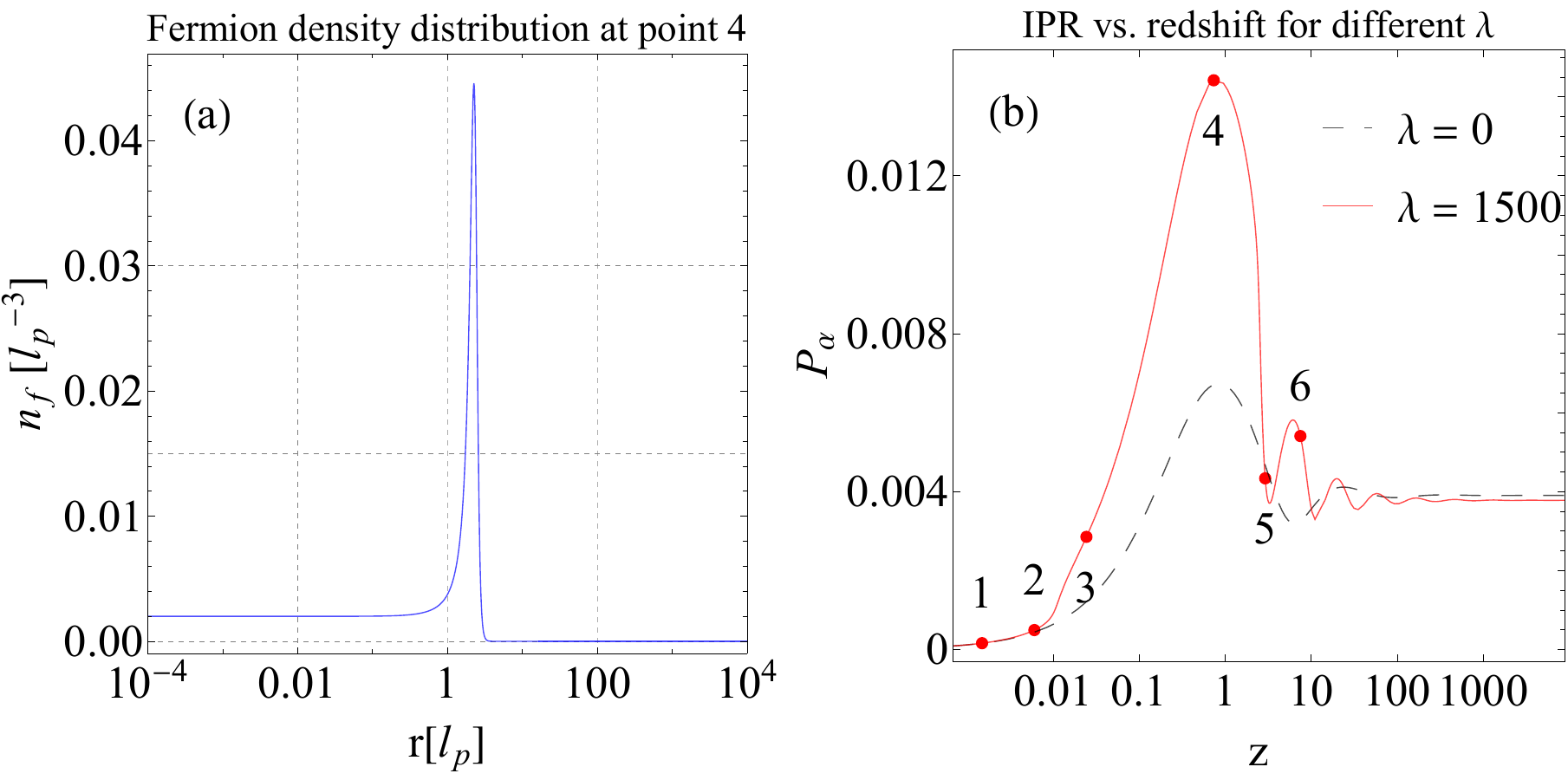}
\caption{\label{fig:fig_massradius} (a) The fermion density at point $4$. (b) The IPR of $\alpha(r)$ as a function of central redshift $z$ for $\lambda=0$ and $\lambda=1500$. The IPR of $\alpha(r)$ for $\lambda=1500$ increases significantly after point $2$ compared to that of the noninteracting case.}
\label{fig:IPR}
\end{figure}

It is interesting to study the behavior of wavefunctions around the new trough of the curve as the wavefunctions continuously evolve. We take the case of $\lambda=1500$ as an example and show in Fig. \ref{fig:spectra}(c1)-(c6) the distributions of wavefunctions $\alpha(r)$ and $\beta(r)$ corresponding to points $1$ to $5$ in the Fig. \ref{fig:spectra}(b). We can see that the amplitude of wavefunction $\beta(r)\to0$ at the low central redshift region (point $1$), where both $\alpha(r)$ and $\beta$ have a larger width. The amplitude of these functions starts to grow as the redshift increases. At the same time, the width of the functions continues to decrease. The effect gets more significant after entering the first trough (points $2$ and $3$). The peak of the function $\alpha(r)$ reaches maximum at point $4$, where the function $\beta(r)$ almost overlaps with $\alpha(r)$ except at the position $r<1$. The fermionic field density corresponding to point $4$ is plotted in Fig. \ref{fig:IPR}(a), which shows high density in the outer radius. Recalling that the solutions are spherically symmetric, this implies a dense fermion outer shell that will become thinner after going through the first trough of the curve. A new peak in wavefunctions appears after leaving the second sharp trough of the curve (points $5$ and $6$), as shown in Fig. \ref{fig:spectra} (c5) and (c6). We define the inverse participation ration (IPR) $P_\alpha$ to characterize the extent of localization for the function $\alpha(r)$,
\begin{equation}
    P_\alpha=\int_0^{+\infty}\alpha(r)^4{\rm dr}\bigg/\int_0^{+\infty}\alpha(r)^2{\rm dr}.
\end{equation}
$P_\alpha=1/L$ denotes the fully delocalized state, where $L$ is the size of the system. A value of $P_\alpha$ close to one corresponds to a highly localized state. We plot the IPR of $\alpha(r)$ as the function of central redshift in Fig. \ref{fig:IPR}(b). We can see that the IPR of function $\alpha(r)$ is significantly increased for $\lambda=1500$ after point $2$ compared to that of the noninteracting case and reaches the maximum at point $4$. After that, the IPR drops and goes through some small oscillations to the level of the noninteracting case. 

\begin{figure}[bpt]
\centering
\includegraphics[width=\columnwidth]{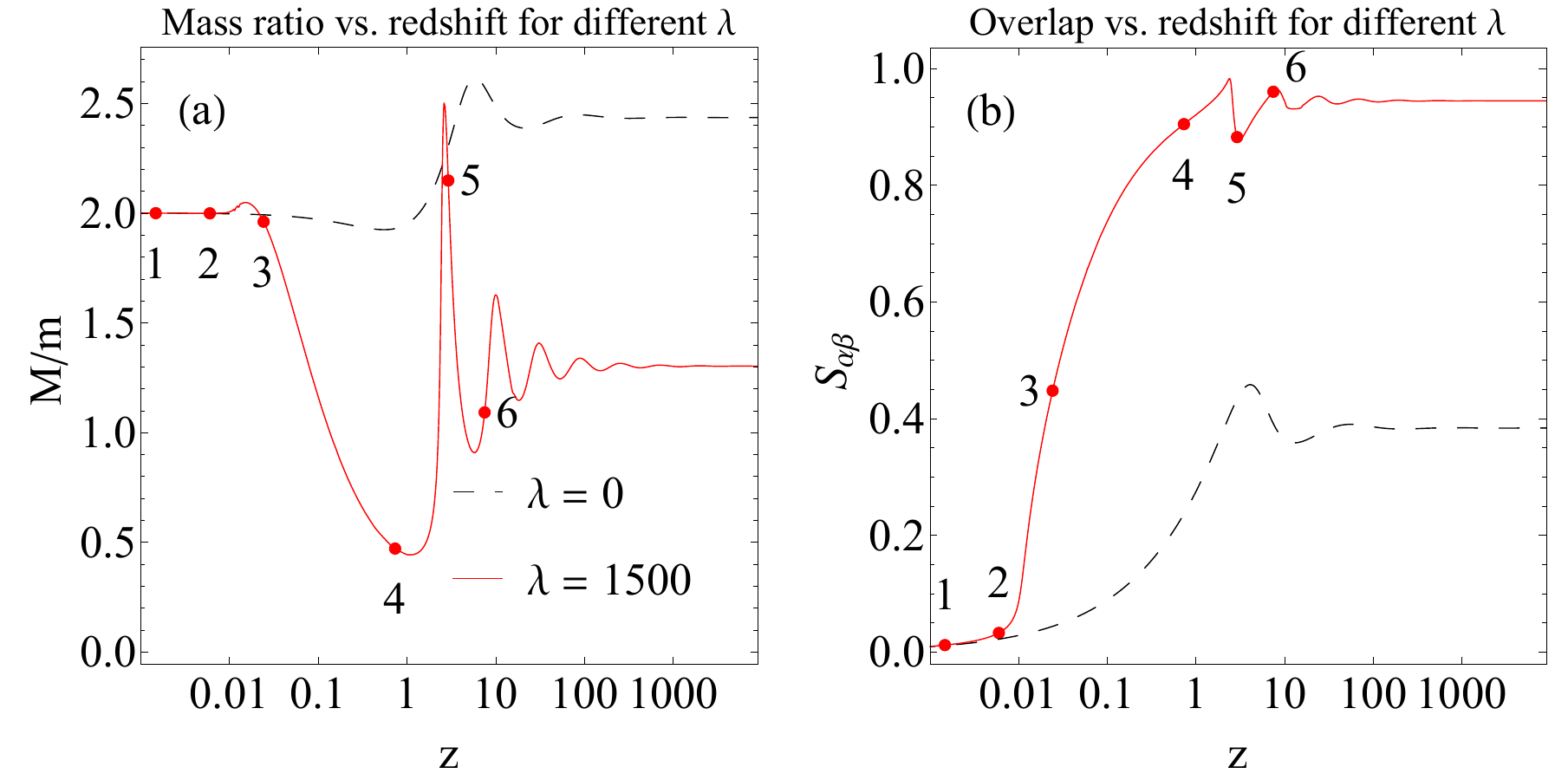}
\caption{\label{fig:fig_massradius} (a) The ratio between ADM mass $M$ and the fermion mass $m$ as a function of central redshift $z$ for different interaction strength $\lambda$. (b) The overlap integral $S_{\alpha\beta}$ of functions $\alpha(r)$ and $\beta(r)$ as a function of central redshift $z$ for different interaction strengths $\lambda$.}
\label{fig:mass_scale}
\end{figure}

From Fig. \ref{fig:spectra}(c4)-(c6), we observe the significant overlap between functions $\alpha(r)$ and $\beta(r)$. If $\alpha(r)\simeq\beta(r)$, the second term of Komar mass \eqref{komar} can be close to zero. Therefore, the change of fermion mass can barely affect the value of Komar mass, leading to the mass-scale separation. This means that the ratio between ADM mass (or equivalently, Komar mass) and fermion mass $M/m$ can be significantly lower than the fermion number $N$. The mass-scale separation has been investigated in Ref. \cite{Leith}, where the authors attribute the reason to the sign switching of the Higgs field. Here, We can define the overlap integral $S_{\alpha\beta}$ to characterize the overlap between two functions
\begin{equation}
    S_{\alpha\beta}=\int_0^{+\infty}\alpha(r)\beta(r){\rm d}r\bigg/\int_0^{+\infty}\alpha^2(r){\rm d}r.
\end{equation}
The more overlap between the two functions, the closer the value of $S_{\alpha\beta}$ is to one. We first plot the ratio $M/m$ as the function of redshift for $\lambda=0,1500$ in Fig. \ref{fig:mass_scale}(a). The mass ratio as $\lambda=1500$ is close to the noninteracting casee in the low redshift regime (points 1 and 2). It drops significantly after point 3 and reaches the minimum $M/m\approx0.4427 $ around point 4. After point 4, the mass ratio grows to a large value close to that of the noninteracting case (point 5) and then goes through damped oscillations to the approximate equilibrium value of $M/m\approx1.3022$. This can be explained by the large change rate of fermion mass in this region, which will be discussed in the next paragraph. We plot the overlap integral $S_{\alpha\beta}$ as a function of redshift in Fig. \ref{fig:mass_scale}(b). We can see that the value of $S_{\alpha\beta}$ starts to grow around point 2 but not around the expected point 3. This could be due to the slow growth of the fermion mass in the low redshift region. The value of fermion mass at points 2, 3, and 4 are $0.2145,\,0.1422,\,1.8033$, respectively, confirming our speculation. The value of $S_{\alpha\beta}$ continues growing after point 3 and reaches a value close to one (points 4, 5, and 6). The corresponding mass ratio after point 4 experiences large damping oscillations. 
\begin{figure*}
\centering
\includegraphics[width=2\columnwidth]{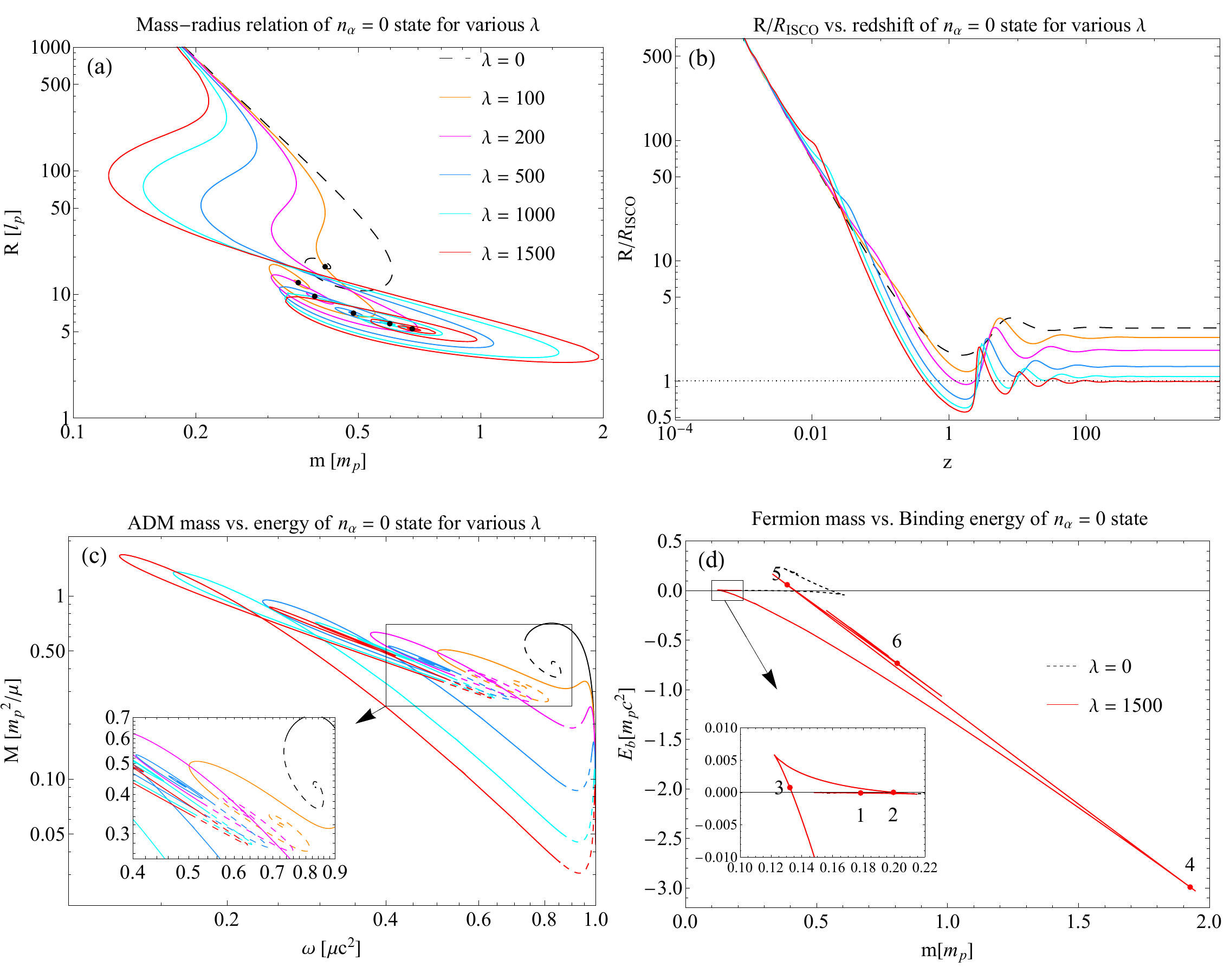}
\caption{ (a) The fermion mass $m$ of the ground state as a function of central redshift $z$ for different interaction strength $\lambda$. (b) The ratio between fermion radius $R$ and $R_{\rm ISCO}$ for various $\lambda$. (c) The ADM mass $M$ as the fermion energy $\omega$ for various $\lambda$ using the same units as in Ref. \cite{Dzhunushaliev3}. The dashed parts of the curve represent the unstable regions. The black line represents the case of $\lambda=0$. Here, $\mu$ is the fermion mass in SI. It differs from $m$, which is in the unit of Planck mass $m_p$. (d) The fermion binding energy $E_b$ as the function of the fermion mass $m$. The six red points on the curve correspond to the points in Fig. \ref{fig:spectra}(b).}
\label{fig:massradius}
\end{figure*}

\begin{figure}[bpt]
\includegraphics[width=\columnwidth]{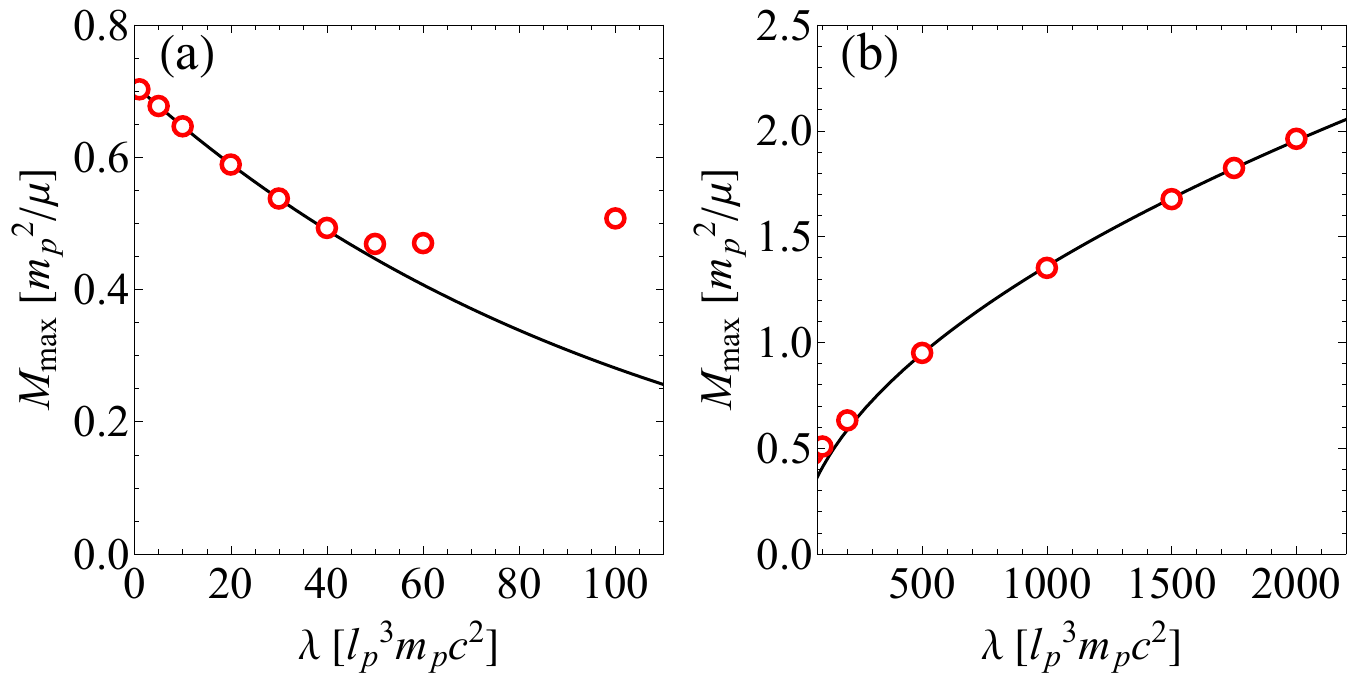}
\caption{(a)-(b) ADM mass $M_{\rm max}$ as a function of the strength of fermionic attractive self- interaction $\lambda$. The red open circles represent the numerical results. The black solid line in (a) ((b)) represents the fitting relation $M_{\rm max}=0.709223\,\exp(-0.00926009\lambda)\,m_p^2/\mu$ ($M_{\rm max}=0.0367514\,\lambda^{0.522738}\,m_p^2/\mu$) for small (large) values of $\lambda$.}
\label{fig:m_max_scaling}
\end{figure}

\begin{figure*}[bpt]
\centering
\includegraphics[width=2\columnwidth]{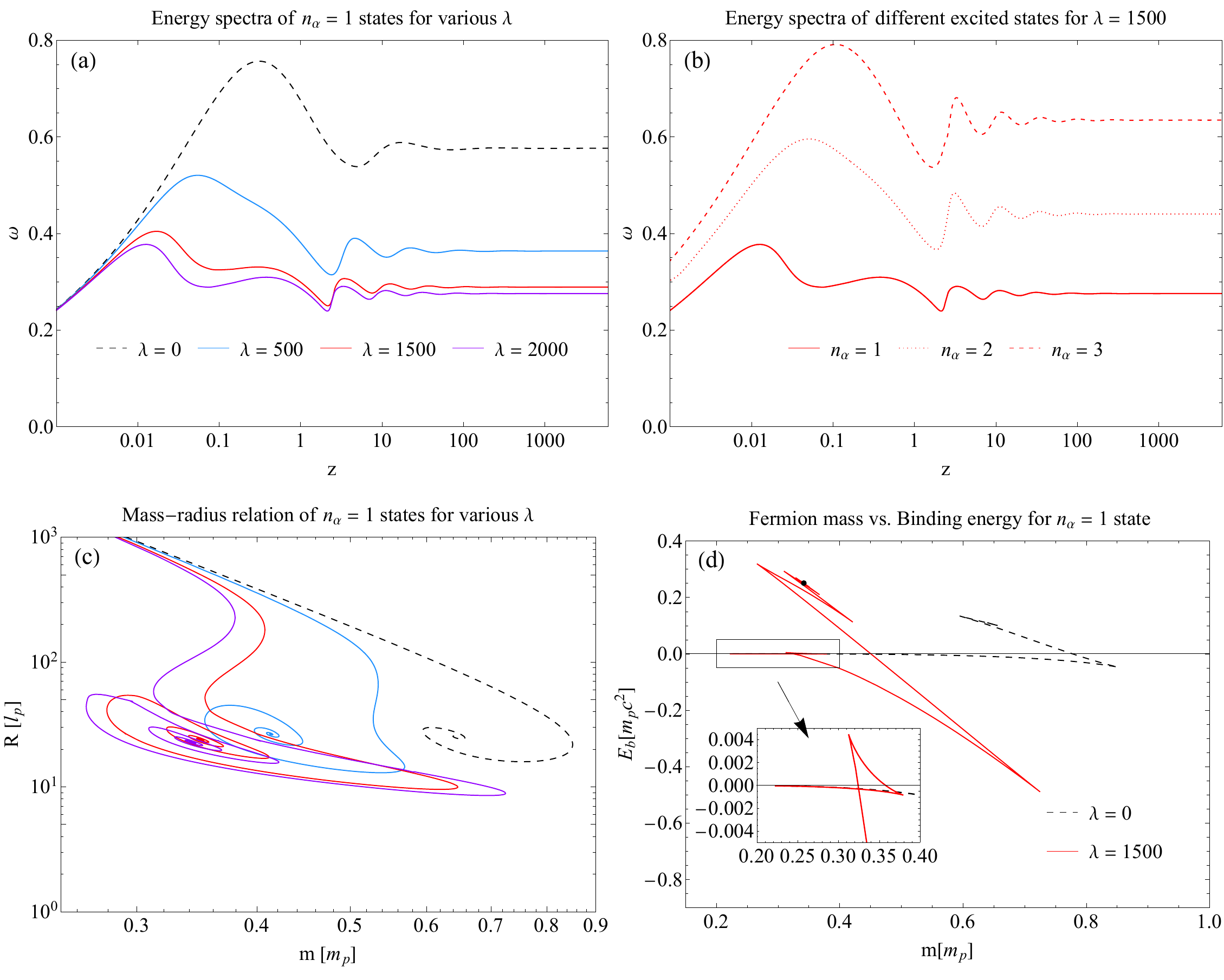}
\caption{ (a) The fermion energy $\omega$ of the first excited state as the function of redshift $z$ for different $\lambda$. The black point on the curve denotes the infinite redshift solution (we use the solution for $z>2\times10^5$ as approximations) (b) The fermion energy $\omega$ as a function of redshift $z$ for different excited states. (c) The mass-radius relations for various $\lambda$ for the first excited states. (d) The fermion binding energy $E_b$ of the first excited states as a function of the fermion mass $m$. The inset shows a small unstable region in the lower redshift regions.}
\label{fig:es_spectra}
\end{figure*}

We plot the mass-radius relations of fermions for different $\lambda$ corresponding to Fig. \ref{fig:spectra}(a) in Fig. \ref{fig:massradius}(a). We can see that the curve takes two more turns before entering the spiral. The mass-radius spiral in high redshift regions becomes much more flattened, resulting in drastic oscillations of the fermion mass. The maximum fermion mass can become larger than the Planck mass $m_p$ for strong enough attractive self-interaction. In contrast, the fermion mass for a noninteracting Dirac star never surpasses the Planck mass. At low redshift region, the mass-radius relation follows the scaling law $m\propto R^{-1/3}$ \cite{Leith3} as in the noninteracting case. The deviation from the scaling law becomes larger as the strength of the interaction increases. The black points in Fig. \ref{fig:massradius}(a) represent the infinite central redshift limit of each mass-radius spiral (The infinite redshift power law solutions are discussed in Appendix.\ref{appendix:inif_z}). The strength of the attractive self-interaction cannot be arbitrarily strong since the system will become unstable and collapse into a black hole if the radius $R$ of fermions defined by Eq. \eqref{radius} becomes smaller than the radius  $R_{\rm ISCO}=6M$ of the innermost stable circular Keplerian orbit (ISCO). Fig. \ref{fig:massradius}(b) shows the ratio between fermion radius and radius of ISCO as a function of central redshift. As we can see, the stronger self-interaction will squeeze the fermions into the region with a smaller radius of only a few Planck lengths. The dashed critical line $R=R_{\rm ISCO}$ intersects with the curves except for the case of $\lambda=0$ and $100$. Additionally, the limit of $R$ for large central redshift solutions is slightly below the critical line. This tells us that the bounded soliton solution for strong attractive interaction is unstable under perturbations and vulnerable to collapse.

In Ref. \cite{Herdeiro}, the authors obtained the maximum ADM mass
\begin{equation}\label{m_max}
M_{\rm max}\approx0.709\,m_p^2/\mu
\end{equation}
for Dirac stars supported by noninteracting fermionic fields, where $\mu$ is the fermion mass in the SI unit. Ref. \cite{Dzhunushaliev3} obtained a scaling law of the maximum of ADM mass
\begin{equation}\label{m_max2}
M_{\rm max}\sim0.415\,\sqrt{|\bar\lambda|}\,m_p^2/\mu
\end{equation}
for Dirac stars supported by repulsive ($\bar\lambda<0$) interacting fermionic fields when $|\bar\lambda|\gg0$, ignoring the case of attractive interaction for the reason that the maximum ADM mass will decrease. To compare to the results in Ref. \cite{Dzhunushaliev3}, we plot the ADM mass $M$ as the function of fermion energy $\omega$ in Fig. \ref{fig:massradius}(c), adopting the same units\footnote{Although the unit of $\lambda\,[l_p^3m_pc^2]$ in this paper is different from that of $\bar\lambda\,[4\pi\hbar^3/m_p^2c]$ in Ref. \cite{Dzhunushaliev3}, it is easy to verify the simple relation $\bar\lambda=\lambda/4\pi$, implying they are in the same order.} as in Ref. \cite{Dzhunushaliev3}. The black (colored) line in Fig. \ref{fig:massradius} (c) represents the energy-mass spiral of the noninteracting (interacting) self-gravitating fermionic field. We confirm the relation \eqref{m_max} in the Ref. \cite{Herdeiro}. We can see that the energy-mass curve for $\lambda>0$ can intersect with itself for strong enough attractive self-interaction, which is similar to the result in Fig. 5 of Ref. \cite{Herdeiro3}. While here we find that $M_{\rm max}$ increases as $\lambda$ becomes larger and can exceed the limit of $0.709\,m_p^2/\mu$ for strong enough attractive interaction ($M_{\rm max}\approx0.950\,m_p^2/\mu,\,1.352\,m_p^2/\mu,\,1.678\,m_p^2/\mu$ for $\lambda=500,\,1000,\,1500$, respectively). However, the radii $R$ for all three solutions that reach maximum ADM mass are less than the corresponding $R_{\rm ISCO}$, implying possible instability. It should be emphasized here that the curves in Fig. \ref{fig:massradius}(c) represent families of solutions with varying fermion mass $\mu$ and the fixed $U(1)$ Noether charge $N$. Those same curves can also be seen as families of solutions with fixed fermion mass $\mu$ and the varying $U(1)$ Noether charge $N$ (with unit $m_p^2/\mu^2$) \cite{Herdeiro2}. Thus, the maximum ADM mass of self-gravitating nonlinear Dirac fields can exceed the limit for a noninteracting Dirac star in the classical sense. We also observe that for small values of $\lambda$, $M_{\rm max}$ decreases as $\lambda$ increases, indicating the existence of minimum of $M_{\rm max}$ for small $\lambda$, as shown in Fig. \ref{fig:m_max_scaling}. In this case, $M_{\rm max}$ approximately follows an exponential scaling 
\begin{align}
    M_{\rm max}& \sim0.709223\,\exp(-0.00926009\lambda)\,m_p^2/\mu\\ \nonumber
    & \approx 0.709223\,\exp(-0.116377\bar\lambda)\,m_p^2/\mu
\end{align}
For large value of $\lambda$, a power law scaling similar to \eqref{m_max2} approximately applies
\begin{align}
    M_{\rm max}& \sim0.0367514\,\lambda^{0.522738}\,m_p^2/\mu\\ \nonumber
    & \approx0.137998\,{\bar\lambda}^{0.522738}\,m_p^2/\mu,
\end{align}
but with a smaller prefactor. The dashed parts of the curve in Fig. \ref{fig:massradius}(c) denote the unstable regions where the binding energy $E_b=M-Nm$ is positive. We can see that there are two unstable regions in the energy-mass spiral for $\lambda>0$. The first unstable region appears after the first maximum of the spiral but does not extend to the infinite central redshift limit, contrary to the noninteracting case. The second unstable region, which is also finite, resides in the highly dense spiral, as shown in the inset. We plot the binding energy $E_b$ as a function of fermion mass $m$ in Fig. \ref{fig:massradius} (d) for $\lambda=1500$. The six red points in the figure correspond to the points in Fig. \ref{fig:spectra} (b). We can see that a large portion of the curve is negative. Unlike the results in Ref. \cite{Leith} where the bound states exist for strong enough coupling at arbitrary central redshift, there is a small region showing unbound solutions at the lower redshift region in our system, as shown in the inset of Fig. \ref{fig:massradius} (d). The positive binding energy is only a necessary condition for stable solutions. A special treatment of linear analysis similar to Ref. \cite{Finster} should be applied here, but it is outside the scope of this paper.

\subsection{$N>2$ and $n_\alpha>0$}
The effect of self-interaction on the excited states only has quantitative differences with that of the ground states shown in Fig. \ref{fig:spectra}. Fig. \ref{fig:es_spectra} displays energy spectra of the first excited states for different $\lambda$. We can see that the curve only gets slightly distorted and does not form the trough as in the ground state when $\lambda=500$. A new trough of the curve forms when $\lambda=1500$ and $2000$. The unscaled excited energy spectra below $\lambda=1500$ do not exhibit multivaluedness, unlike the ground state spectra, indicating a much larger threshold for the emergence of multivaluedness. Higher excited states also show a much slighter distortion of the energy curves at the first trough amid the sharpening around the other maxima and minima of the curve, as shown in Fig. \ref{fig:es_spectra}. Fig. \ref{fig:es_spectra}(c) displays the mass-radius relation of the first excited states corresponding to Fig. \ref{fig:es_spectra}(a). It can be seen that the excited state has a larger minimum of the radius and a smaller maximum of the fermion mass compared to those of the ground state for the same value of $\lambda$. Fig. \ref{fig:es_spectra} shows the binding energy of the excited state as a function of the fermion mass. It also shows a small, unstable region for the lower central redshift. Whereas the infinite redshift solution is unstable in this case since its binding energy is positive.

\begin{figure}
\centering
\includegraphics[width=\columnwidth]{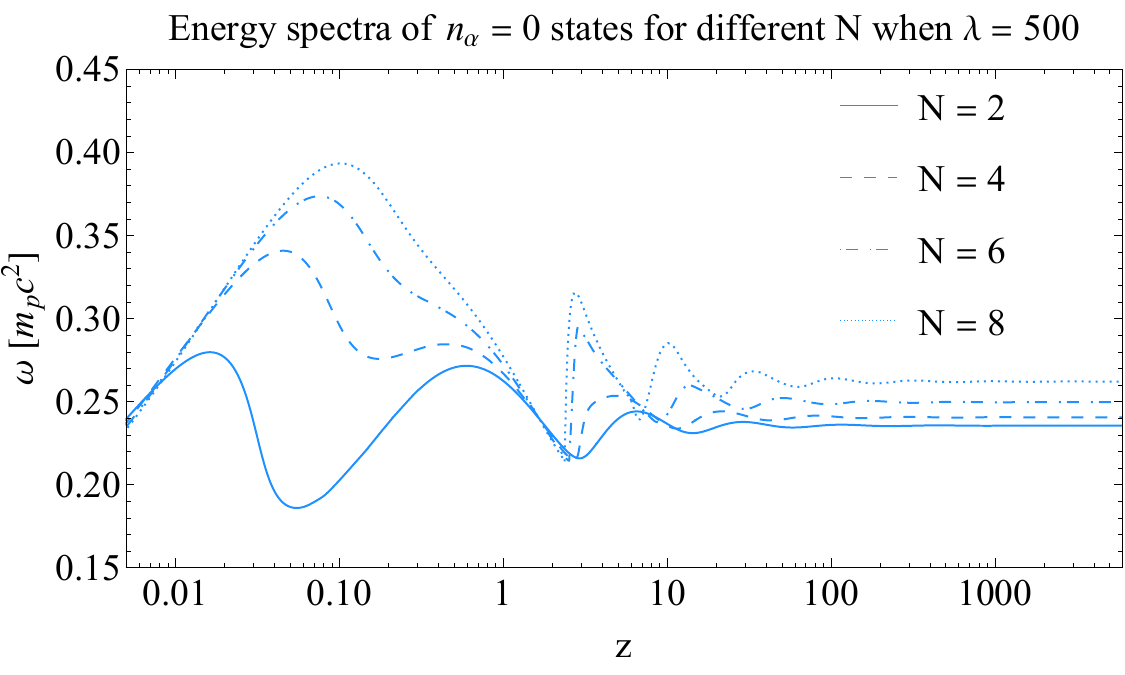}
\caption{The ground state energy spectra for different fermion number N when $\lambda=500$.}
\label{fig:N_spectra}
\end{figure}

\begin{figure*}
\centering
\includegraphics[width=2\columnwidth]{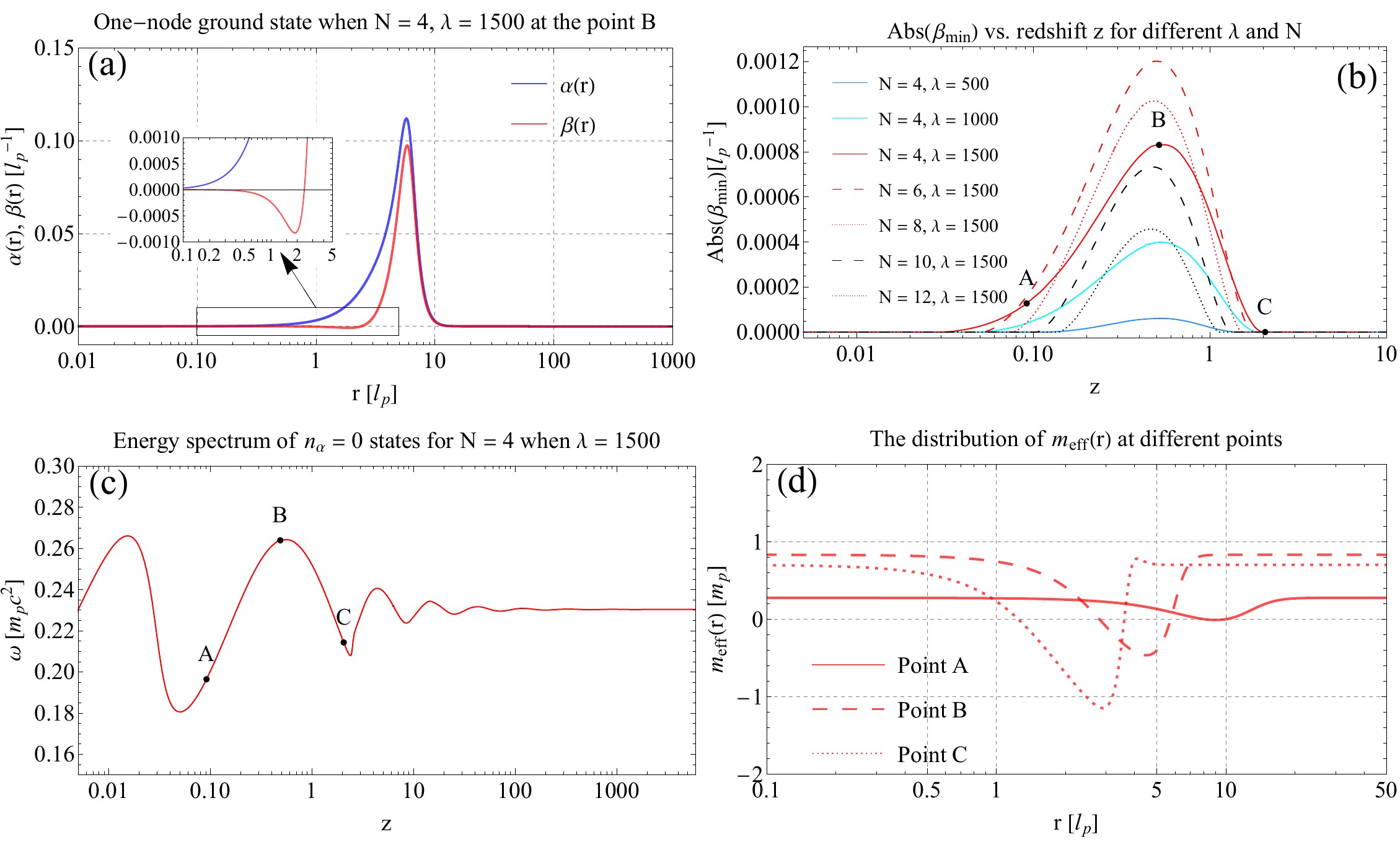}
\caption{(a) The odd parity ground state solution obtained from wavefunction ansatz \eqref{ansatz}. The inset in the figure clearly shows the existence of the node in the function $\beta(r)$. The corresponding parameters are $(z,m,\omega\approx0.4895,\,0.8333,\,0.2639)$. (b) ${\rm Abs}(\beta_{\rm min})$ as a function of the central redshift $z$ for different fermion number $N$ and strengths of interaction $\lambda$. Point A denotes the position where the effective mass $m_{\rm eff}(r)$ starts to possess negative values. Point B represents the maximum of the solid red line. Point C denotes the transition of solution with different parities in high redshift regions. (c) The ground state energy spectrum for $N=4$ and $\lambda=1500$. (d) The distribution of $m_{\rm eff}$ at different points.}
\label{fig:betamin}
\end{figure*}

For the same strength of self-interaction, adding more fermions to this system can also reduce the multivaluedness and the distortion of the energy spectra at the lower redshift, as shown in Fig. \ref{fig:N_spectra} (In Fig. \ref{fig:N_spectra}, only the case $N=2$ exhibits multivaluedness for unscaled energy spectra). We can see that the first trough after the first maximum almost disappears when $N=8$, which is similar to the results in Fig. \ref{fig:es_spectra}(b). Meanwhile, troughs after that become sharper if the fermion number $N$ increases, as in the noninteracting case at the higher redshift regions. From previous discussion, we know that the appearance of the first trough in the energy spectrum is related to the strength of fermion self-interaction. The similarity of the behaviors at the lower redshift regions shown in Fig. \ref{fig:es_spectra}(b) and Fig. \ref{fig:N_spectra} can be explained by the quantum-originated pressure that counterbalances the self-attraction between fermions. At the lower redshift regions where the gravity is weak, solutions at higher excited states are more delocalized, and solutions that include more fermions have larger quantum degenerate pressure due to the Pauli exclusion principle. Although our fermion wavefunctions are semi-classical, we believe the same effects can be found for a fully quantized fermionic field \cite{Kain}.

\begin{figure*}
\centering
\includegraphics[width=2\columnwidth]{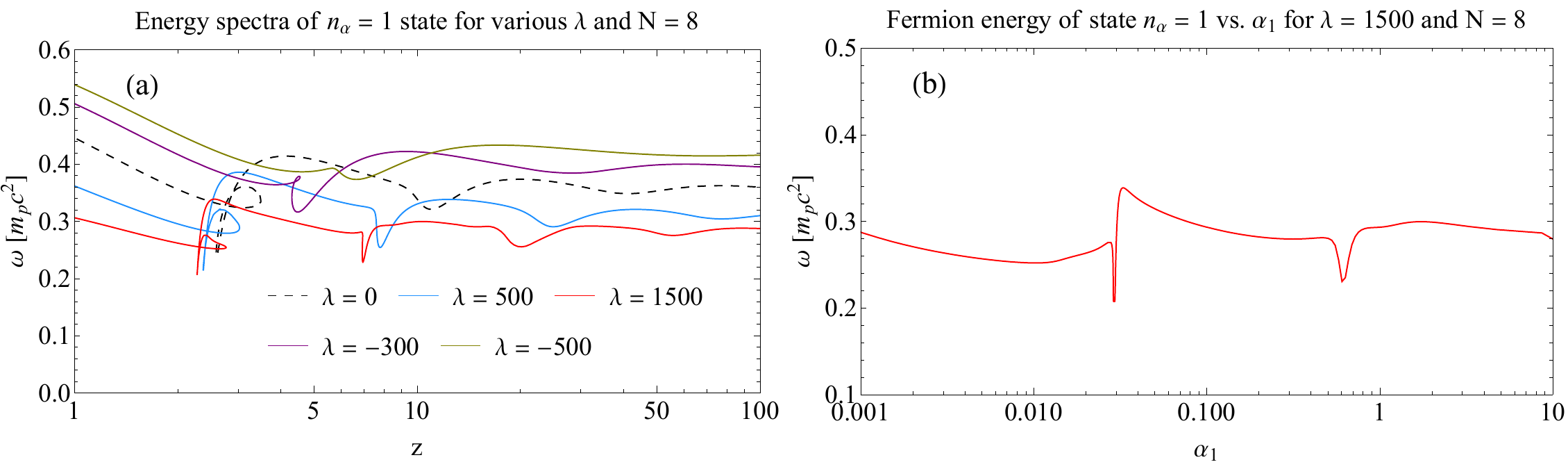}
\caption{(a) The first excited state energy spectra for different $\lambda$ when $N=8$ in the high redshift region. (b) The first excited state energy as a function of unscaled parameter $\alpha_1$ when $N=8$ and $\lambda=1500$. The curve is single-valued.}
\label{fig:es_spectra2}
\end{figure*}

In Sec. \ref{sec:IV}, we discussed the reason for using the node number $n_\alpha$ of function $\alpha(r)$ to label different families of states. The evidence of the existence of the one-node ground state is shown in Fig. \ref{fig:betamin}(a). The inset in the figure clearly shows the existence of a zero and a negative minimum in the function $\beta(r)$ for $N=4$ and $\lambda=1500$. A solution with different parity can be found by simply setting $m=-1$ in the noninteracting Einstein-Dirac equations as explained in Ref. \cite{Finster}, which leads us to understand this behavior by the effective mass of the fermion
\begin{equation}
    m_{\rm eff}=m-\lambda(\bar\Psi^\pm\Psi^\pm)=m-\frac{N\lambda T(\alpha^2-\beta^2)}{r^2}.
\end{equation}
Then, the original equation of motion for the Dirac field (\ref{eom}a) becomes
\begin{equation}
(\slashed{D}-m_{\rm eff})\Psi^+=0.
\end{equation}
Note that $\Psi^-=\gamma^5\Psi^+$, it is easy to verify that $\Psi^-_{j,k}$ satisfies
\begin{equation}
(\slashed{D}+m_{\rm eff})\Psi^-=0.
\end{equation}
Thus, solutions of negative effective fermion mass can be interpreted as solutions of positive effective fermion mass with a different parity. The effective mass $m_{\rm eff}$ should be negative in local regions for large enough $\lambda$, $N$ and $T(r)$ since $m_{\rm eff}\to m$ if $\alpha,\beta\to0$. Note that we are also dealing with localized solutions of Eqs. \eqref{ede}, it is sufficient to consider the local effective fermion mass. We use the absolute value of function $\beta(r)$'s minimum ${\rm Abs}(\beta_{\min})$ to characterize the effect of the negative fermion mass. We plot ${\rm Abs}(\beta_{\min})$ as a function of redshift $z$ for different fermion numbers $N$ and strength of interaction $\lambda$ in Fig. \ref{fig:betamin}(b). We can see that the values of ${\rm Abs}(\beta_{\min})$ for all cases reach maxima in the low redshift regions and gradually decrease to zero in the relativistic regime ($z>1$) that displays stronger gravity. This tells us that solutions with different parities can smoothly transform into each other as the central redshift increases. The $\beta_{\min}$ becomes more negative for the same $N$ if $\lambda$ increases. However, increasing the fermion number $N$ for the same $\lambda$ does not show this trend. The $N=8$ red dotted line is lower than the $N=6$ red dashed line and even lower than the $N=4$ red solid line in some regions. Further increasing the fermion number can suppress the effect of parity violations. As we can see from {\ref{fig:betamin}}(b), the $N=10$ black dashed line and $N=12$ black dotted line are both lower than the $N=4$ red solid line for the same interaction strength. This might be explained by the much stronger gravitational effect for the system with more fermions. The point A is the location where the effective mass $m_{\rm eff}$ starts to exhibit negative values. Notice that there is still a small region where the solution shows a different parity while the effective mass $m_{\rm eff}$ is positive for all $r$, the reason behind this is unclear. Point B is the maximum of the line $N=4,\lambda=1500$ corresponding to the solution in Fig. \ref{fig:betamin}(a). Point C is the approximate transition point between solutions with different parities in the high redshift region. The location of three points in the energy spectrum is shown in Fig. \ref{fig:betamin}. We can see that point B is sitting around the second maximum of the curve. It should be emphasized here that the effective mass $m_{\rm eff}(r)$ always has negative part after point A since the function $T(r)$ can increase without bound for the same value of $N$ and $\lambda$. Therefore, the effective fermion mass can only partially explain the parity transition of the solutions. We show the distribution of $m_{\rm eff}$ for these points in Fig. \ref{fig:betamin}(d). It can be seen that $m_{\rm eff}$ has a smaller value at point C although the corresponding solution's broken parity symmetry has been restored. We have also verified that there is no parity transition in another branch of solutions found by setting $m=-1$.

From the above discussion, we conclude that the system favors solutions with a different parity when its central redshifts are located at ranges around $z=1$, suggesting a possible parity violation. Whereas the original Eqs. \eqref{eom} are invariant under parity transformation, the system can experience spontaneous parity symmetry breaking by tuning the central redshift. In fact, the solutions with broken parity symmetry can be found for the nonlinear Dirac equation in flat spacetime(See Appendix. \ref{flat_gs}). The coupling with gravitation suppresses the symmetry-breaking effect and restores the original parity at a strong gravitation regime, resulting in the disappearance of node of the ground state(value of ${\rm Abs}(\beta_{\rm min})$ decreasing to zero), as shown in Fig. \ref{fig:betamin}(b).

Finally, we discuss how the interaction can affect the loop structure in excited spectra discovered in Ref. \cite{Leith3}. Fig. \ref{fig:es_spectra2}(a) shows the energy spectra of $n_\alpha=1$ states for different strengths (repulsive or attractive) of interaction when $N=8$. We can see that the loop structure in the excited energy spectrum gradually shrinks and disappears, and the sharp minimum becomes smoother as the repulsive interaction increases compared to the noninteracting case. The $\omega-z$ becomes single-valued in the end. Confirming the conclusion in Ref. \cite{Leith3} that the folding of excited energy spectra is due to the fermion self-trapping effect. In comparison, the results for the case of attractive interaction are subtle. The entire energy spectra shift to the lower energy regime while the multivalued loops also shrink slightly but do not disappear. Surprisingly, we find that the multiplicity of excited state solutions in the high redshift regime disappears when $\lambda=1500$. As a result, $\alpha_1$ becomes an ideal quantity to parameterize the excited spectrum, as shown in Fig. \ref{fig:es_spectra2}(b). The first narrow trough in Fig. \ref{fig:es_spectra2}(b) corresponds to the loop structure of the red line in Fig. \ref{fig:es_spectra2}(a). We also find multiplicity of ground state solutions for $\lambda=1500$ and $N=8$. It shows another hint that the multiplicity of excited solutions around the loop structure could have a different origin from that in the ground state. We speculate that the multiplicity around the loop structure is due to the quantum degenerate pressure between fermions, since adding more fermions in this system can induce a further degree of multiplicity of solutions, while this multiplicity can be reduced by increasing the strength of the attractive self-interaction between fermions.
\section{Conclusion and discussion}\label{sec:VI}

In this paper, we investigate the behaviors of localized semi-classical spherically symmetric solutions for Einstein-Dirac equations with the scalar nonlinearity of spinors and focus on the case of attractive interaction, where the fermionic fields are represented by the first quantized wavefunctions. We find the multiplicity of solutions for ground or excited states in this system for strong enough interaction, similar to that in the Einstein-Dirac-Higgs system \cite{Leith}. The threshold of the interaction strength for the appearance of excited state multiplicity is much higher. This multiplicity of solutions is inherent in the nonlinear Dirac solutions and does not imply the multivaluedness of the corresponding rescaled energy spectrum (Fig. \ref{fig:spectra}(a)-(b)). We suggest the origin of this multiplicity of solutions in our system is different from that in Ref. \cite{Leith3} since we can eliminate this multiplicity of excited states found for fermion numbers larger than 4 by increasing the strength of attractive fermion self-interaction (Fig. \ref{fig:es_spectra2}). 

The curve of fermion energy vs. central redshift can develop a new trough in low redshift regions, indicating the instability of the solutions. The fermion density continuously evolves through this trough and becomes much denser at the outer radius (Fig. \ref{fig:spectra} and Fig. \ref{fig:IPR}). This new trough can gradually disappear by increasing the index of excited states and the fermion number in the system (Fig. \ref{fig:es_spectra} and Fig. \ref{fig:N_spectra}). We also find a similar mass-scale separation between the fermion mass and ADM mass of the solution as in the Einstein-Dirac-Higgs system (Fig. \ref{fig:mass_scale}). In our system, this mass-scale separation can be explained by the increase in overlap between the two spinor components. The maximum ADM mass of solutions in our system can exceed the limit for noninteracting Dirac stars for strong enough fermion self-interaction in the classical sense (Fig. \ref{fig:massradius}). The corresponding radius is less than the innermost circular Keplerian orbit and thus vulnerable to collapse. However, if the strength of attractive self-interaction is small, the maximum ADM mass decreases as self-interaction becomes stronger. The $M_{\rm max}$ exhibits different scaling laws with small and large values of $\lambda$ (Fig. \ref{fig:m_max_scaling}). The nonlinear Dirac equations can show parity violations for strong enough attractive interactions, which can be explained by the negative effective mass of fermions (Fig. \ref{fig:betamin}). The coupling of nonlinear Dirac equations with Einstein's gravity can restore the parity symmetry, behind which the mechanism is unclear. 

In conclusion, we have found some novel behaviors, e.g., parity violation of solutions, of nonlinear quantum matter coupled with Einstein's gravity at the semi-classical level, and clarified some issues (multiplicity of solutions) in the previous works. In future works, it is worth investigating if the Einstein-Dirac-Higgs system, where the fermion mass is produced by the Higgs field, can show similar parity violations. Extending the present system to a fully quantized version is possible \cite{Kain}. Another direction is to consider the ``Dirac stars'' in the modified gravity framework, similar work has been done on the ``boson stars'' \cite{Masó-Ferrando,Masó-Ferrando2}.

\begin{acknowledgments}
The authors thank Keith Horne for the valuable comments. The research is supported by the National Natural Science Foundation of China (Grant No. 12471239 ), and the Guangdong Basic and Applied Basic Research Foundation (Grant No. 2024A1515013106).
\end{acknowledgments}

\appendix

\section{Infinite redshift power low solutions}\label{appendix:inif_z}
\begin{figure*}[bpt]
\centering
\includegraphics[width=2\columnwidth]{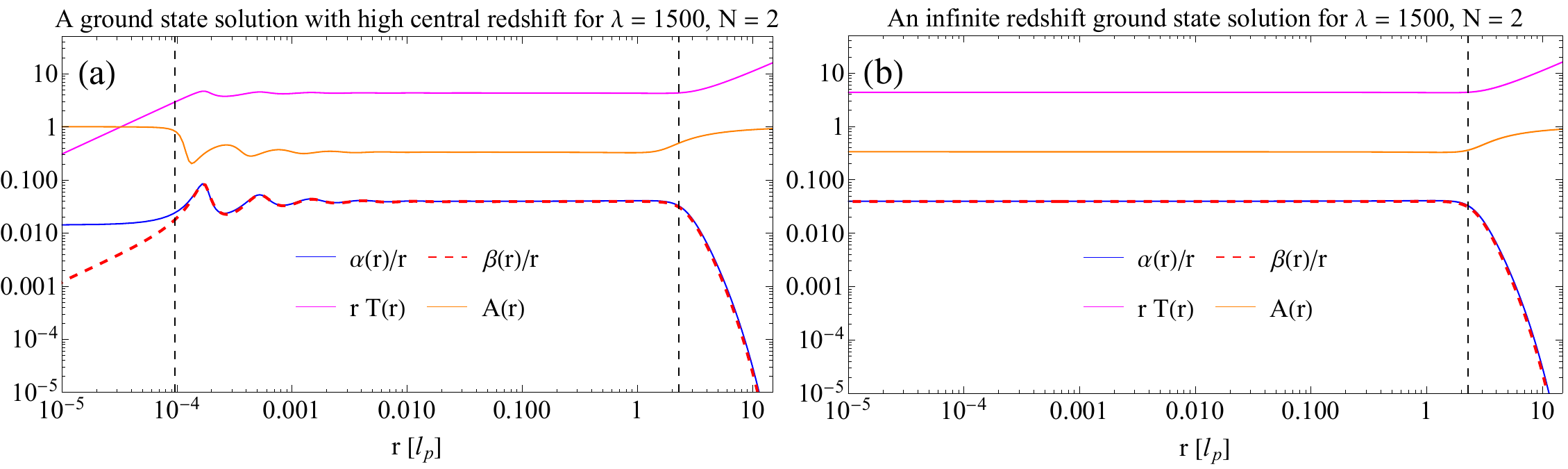}
\caption{(a) The ground state solution of Eqs. \eqref{ede} for large central redshift $z\approx30468.5$ when $\lambda=1500$ and $N=2$, with $m\approx0.679346$ and $\omega\approx0.229131$. The right (left) dashed black vertical line is the radius of $r=R_{1(2)}$. The wave-like zone near $r=R_2$ is a typical sign of self-trapping of fermions. The functions $\alpha(r)$ and $\beta(r)$ are highly overlapped in the power-law zone. (b) The corresponding infinite redshift ground state solution, with $m\approx0.679302$ and $\omega\approx0.229131$. }
\label{fig:powerlaw_sol}
\end{figure*}

\begin{figure}[bpt]
\centering
\includegraphics[width=\columnwidth]{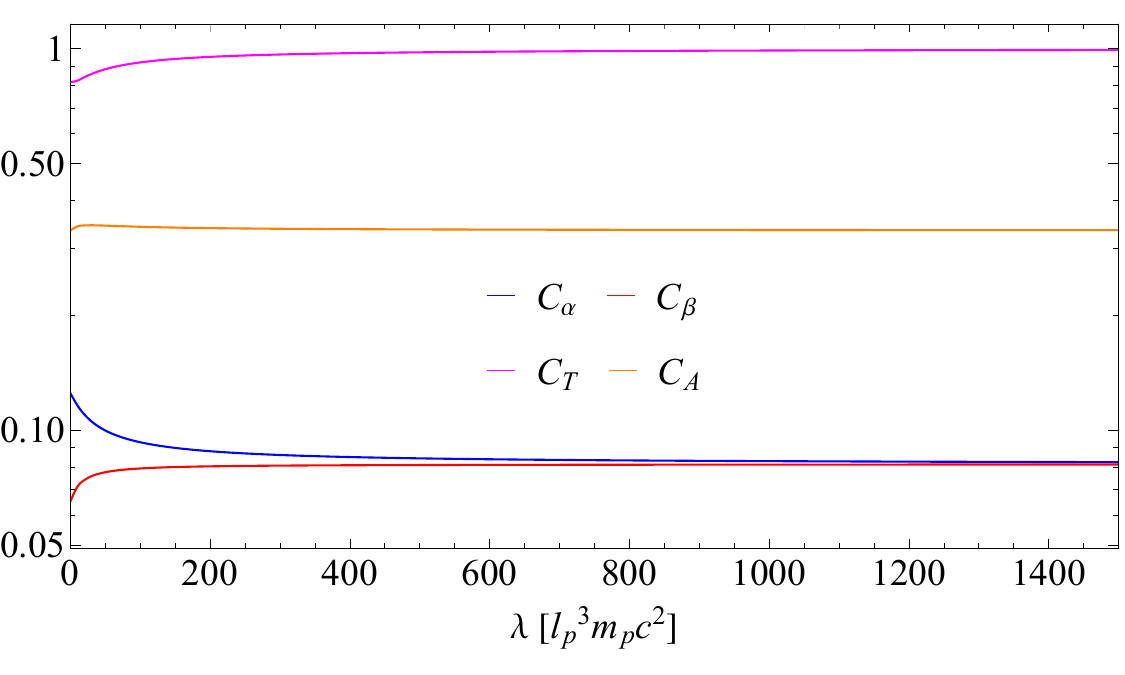}
\caption{The coeffecients $C_\alpha,\,C_\beta,\,C_A$, and $C_T$ as a function of $\lambda$ when $N=2$. }
\label{fig:powerlaw_coeff}
\end{figure}
The ground state structure of the Einstein-Dirac equations with high central redshift can be classified into three different zones \cite{Bakucz}. The first one is the ``core'' zone located at the center. The second one is called the ``power-law'' zone that is continuously connected to the central core zone. The energy $\omega T(r)\gg m$ in the power-law region. There is also an ``evanescent” zone in the outer radius of the solution, where the wavefunctions $\alpha(r)$ and $\beta(r)$ decay exponentially and $\omega T(r)\ll m$. The boundary between the power-law zone and the evanescent zone is where $\omega T(R)\approx m$. One can find the power-law solution by the singular expansion of massless Einstein-Dirac equations. Here, we can also find a similar structure in our system's ground states and the power-law solution, while there is an extra ``wave-like" zone between the core zone and the power-law zone characterizing the self-trapping of fermions. The massless version of Eqs. \eqref{ede} are as follows
\begin{subequations}\label{massless_ede}
    \begin{eqnarray}
        \sqrt{A}\alpha^\prime & = & \frac{N\alpha}{2r}-\Big(\omega T-\lambda V\Big)\beta,\\
        \sqrt{A}\beta^\prime & = & \Big(\omega T+\lambda V\Big)\alpha-\frac{N\beta}{2r},\\
        rA^\prime & = & 1-A-8\pi N\omega T^2(\alpha^2+\beta^2)\nonumber\\
        & & -4\pi r^2\lambda V^2,\\
        2rA\frac{T^\prime}{T} & = & A-1-8\pi NT\sqrt{A}(\alpha\beta^\prime-\beta\alpha^\prime)\nonumber\\
        & & +4\pi r^2\lambda V^2,\\
        V & = & \frac{NT}{r^2}(\alpha^2-\beta^2).
    \end{eqnarray}
\end{subequations}
We seek the power-law solution of the following form
\begin{equation}\label{powerlaw}
    \alpha(r)=\alpha_pr,\,\beta(r)=\beta_pr,\,T(r)=\frac{T_p}{r},\,A(r)=A_p,
\end{equation}
Where $\alpha_p,\,\beta_p,\,T_p$, and $A_p$ are coefficients to be determined. Substituting the ansatz \eqref{powerlaw} into Eqs. \eqref{massless_ede}, we obtain the following constraints for the four coefficients.
\begin{align}\label{constraints}                                                                      
\sqrt{A_p} \alpha_p - \frac{N \alpha_p}{2} - N T_p \beta_p \lambda (\alpha_p^2 - \beta_p^2) + T_p \beta_p \omega & = 0 \nonumber\\
\sqrt{A_p} \beta_p + \frac{N \beta_p}{2} - N T_p \alpha_p \lambda (\alpha_p^2 - \beta_p^2) - T_p \alpha_p \omega & = 0 \nonumber\\
-1 + A_p + 4 \pi N^2 T_p^2 \lambda (\alpha_p^2 - \beta_p^2)^2 + 8\pi N T_p^2 \omega(\alpha_p^2+\beta_p^2) & = 0 \nonumber\\
-1 + 3 A_p - 4 \pi N^2 T_p^2 \lambda (\alpha_p^2 - \beta_p^2)^2 & = 0.
\end{align}
Solutions of $\alpha_p,\,\beta_p,\,T_p$, and $A_p$ for $\lambda=0$ are
\begin{align}\label{coeff_noninteracting}
    \alpha_p & =\frac{3^{1/4}\omega^{1/2}}{\sqrt{6\pi N^2(\sqrt{3}N-2)}},\nonumber\\
    \beta_p & =\frac{3^{1/4}\omega^{1/2}}{\sqrt{6\pi N^2(\sqrt{3}N+2)}},\nonumber\\
    A_p & =\frac{1}{3},\nonumber\\
    T_p & =\sqrt{\frac{N^2}{4}-\frac{1}{3}}\,\omega^{-1}.
\end{align}
We cannot find the explicit expressions for these coefficients when $\lambda>0$. However, we can prove that these coefficients follow the same scaling law with $\omega$ as the relations in \eqref{coeff_noninteracting}
\begin{align}\label{coeff_C}
    \alpha_p & = C_\alpha(\lambda,N)\omega^{1/2},\nonumber\\
    \beta_p & = C_\beta(\lambda,N)\omega^{1/2},\nonumber\\
    A_p & = C_A(\lambda,N),\nonumber\\
    T_p & = C_T(\lambda,N)\omega^{-1}.
\end{align}
Substituting Eqs. \eqref{coeff_C} into Eqs. \eqref{constraints}, we can obtain compact equations for $C_\alpha,\,C_\beta,\,C_A$, and $C_T$ 
\begin{align}
    \sqrt{C_A} C_\alpha - \frac{N C_\alpha}{2} - N C_T C_\beta \lambda (C_\alpha^2 - C_\beta^2) +C_TC_\beta= 0\nonumber \\
    \sqrt{C_A} C_\beta + \frac{N C_\beta}{2} - N C_T C_\alpha \lambda (C_\alpha^2 - C_\beta^2)-C_TC_\alpha = 0 \nonumber\\
    -1 + C_A + 4 \pi N^2 C_T^2 \lambda (C_\alpha^2 - C_\beta^2)^2+8\pi NC_T^2(C_\alpha^2+C_\beta^2) = 0 \nonumber\\
    -1 + 3 C_A - 4 \pi N^2 C_T^2 \lambda (C_\alpha^2 - C_\beta^2)^2 = 0,
\end{align}
which we solved numerically. For $\lambda=1500$ and $N=2$, $C_\alpha\approx0.0824542$, $C_\beta\approx0.0812967$, $C_T\approx0.991926$, and $C_A\approx0.334222$. Now, we can determine the boundary $r=R_1$ between the power-law zone and the evanescent zone in our system by the relation $\omega\, T(r)=\lambda V+m$. Substituting the power-law solutions (Combining Eqs. \eqref{powerlaw}) and Eqs .\eqref{coeff_C}, we can obtain the radius
\begin{equation}
    R_1=[1+N\lambda (C_\alpha^2-C_\beta^2)]C_T/m.
\end{equation}
The boundary between the core zone and the power-law zone is simply $r=R_2=T_p/(1+z)$.

Fig. \ref{fig:powerlaw_sol}(a) shows the ground state solution for $\lambda=1500$ and $N=2$ with high central redshift $z\approx30468.5$. The $\alpha(r)/r,\,\beta(r)/r,\,r\,T(r)$ and $A(r)$ are approximately constant in the power law zone except near the boundary of $r=R_2$, where there is a wave-like zone. This wave-like zone is a typical structure found in Ref. \cite{Leith2}, signaling the self-trapping of fermions. We can also find a localized solution using the power law solution \eqref{powerlaw} as the initial condition. The result is shown in Fig. \ref{fig:powerlaw_sol}(b), which matches the result in Fig. \ref{fig:powerlaw_sol}(a) pretty well in the power-law zone and evanescent zone. No wave-like structure was observed in this case. We would also like to show how the strength of self-interaction can affect the power-law expansion in small $r$. Fig. \ref{fig:powerlaw_coeff} displays how the coefficients $C_\alpha,\,C_\beta,\,C_A$, and $C_T$ are changing with $\lambda$. We can see that $C_T$ and $C_A$ only increase slightly as $\lambda$ increases, while $C_\alpha$ and $C_\beta$ get close to each other. The close value of $C_\alpha$ and $C_\beta$ can explain the highly overlapped wavefunctions in the power-law zone for strong self-interaction.
\section{The solution of nonlinear Dirac equations in the flat spacetime}\label{flat_gs}
\begin{figure}[hb]
\centering
\includegraphics[width=\columnwidth]{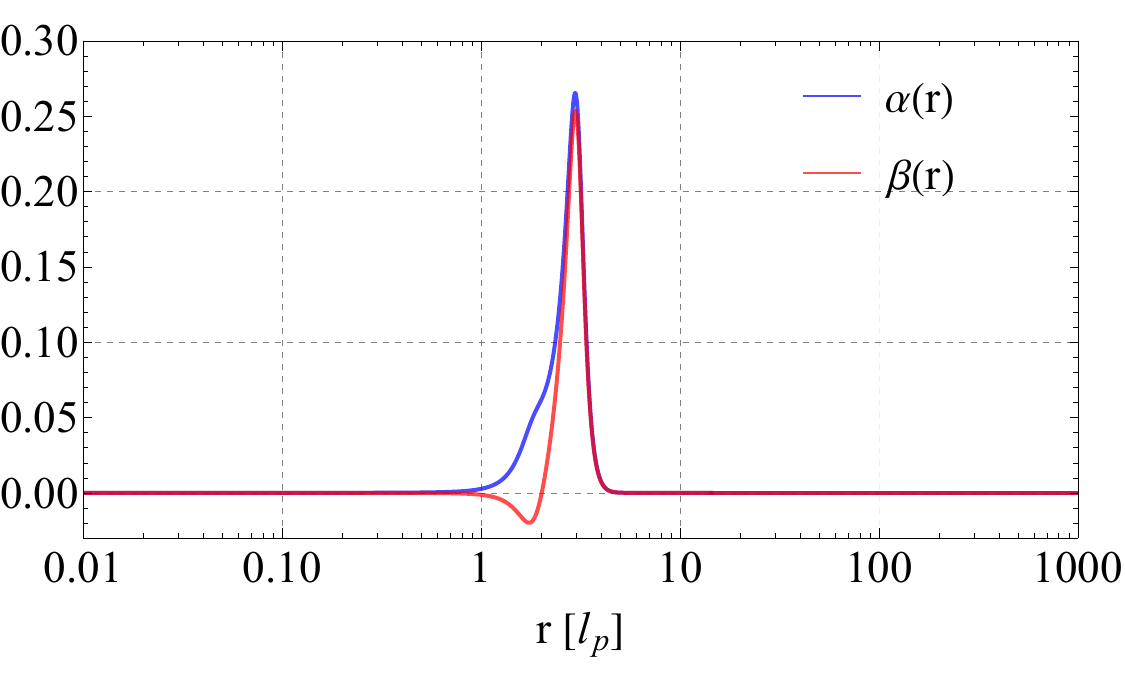}
\caption{The ground state for nonlinear Dirac equations \eqref{flat_nde} in the flat spacetime when $\lambda=1500$ and $N=4$.}
\label{fig:gs_flat}
\end{figure}
The nonlinear Dirac equations in flat spacetime are as follows
\begin{subequations}\label{flat_nde}
    \begin{eqnarray}
       \alpha^\prime & = & \frac{N\alpha}{2r}-\Big(\omega+m-\lambda V\Big)\beta,\\
      \beta^\prime & = & \Big(\omega-m+\lambda V\Big)\alpha-\frac{N\beta}{2r},\\
        V & = & \frac{N}{r^2}(\alpha^2-\beta^2).
    \end{eqnarray}
\end{subequations}

We can find the normalized solution with the constraints
\begin{equation}
    4\pi\int_0^{+\infty}(\alpha^2+\beta^2){\rm d}r=\kappa^2,
\end{equation}
and the rescaled relations
\begin{subequations}
    \begin{eqnarray}
    \tilde{\alpha}(r) & = & \alpha(r)/\sqrt{\kappa},\\
    \tilde{\beta}(r) & = & \beta(r)/\sqrt{\kappa}.
    \end{eqnarray}
\end{subequations}
Fig. \ref{fig:gs_flat} displays one of the solutions for positive fermion mass $m$. We can see one node in the function $\beta(r)$, indicating it is odd parity symmetric.

\end{document}